\documentstyle{mn}
\input{psfig.sty}
\newcommand{\etal}{{et al.~}}

\newcommand{\gta}{\ga}

\newcommand{\kmsmpc}{\>{\rm km}\,{\rm s}^{-1}\,{\rm Mpc}^{-1}}
\newcommand{\kms}{\>{\rm km}\,{\rm s}^{-1}}

\newcommand{\Mpc}{\>{\rm Mpc}}

\newcommand{\Msun}{\>{\rm M_{\odot}}}

\newcommand{\Lsun}{\>{\rm L_{\odot}}}
\newcommand{\MLsun}{\>({\rm M}/{\rm L})_{\odot}}

\newcommand{\beq}{\begin{equation}}
\newcommand{\eeq}{\end{equation}}

\newcommand{\mpch}{\>h^{-1}{\rm {Mpc}}}

\newcommand{\msunh}{\>h^{-1}\rm M_\odot}

\newcommand{\apj}{ApJ}
\newcommand{\apjs}{ApJS}
\newcommand{\aj}{AJ}
\newcommand{\mnras}{MNRAS}

\newdimen\hssize
\newdimen\hdsize 
\begin{document}
            
%%%%%%%%%%%%%%%%%%%%%%%%%%%%%%%%%%%%%%%%%%%%%%%%%%%%%%%%%%%%%%%%%%%%%%%%%%

\title[A Halo-Based Galaxy Group Finder]
      {A Halo-Based Galaxy Group Finder:\\ 
       Calibration and Application to the 2dFGRS}       
\author[Yang, Mo, van den Bosch \& Jing]
       {Xiaohu Yang$^{1}$, H.J. Mo$^{1}$, 
        Frank C. van den Bosch$^{2}$, Y.P. Jing$^{3}$
        \thanks{E-mail: xhyang@astro.umass.edu}\\
      $^1$ Department of Astronomy, University of Massachusetts,
           Amherst MA 01003-9305, USA\\
      $^2$ Department of Physics, Swiss Federal Institute of
           Technology, ETH H\"onggerberg, CH-8093, Zurich,
           Switzerland\\ 
      $^3$Shanghai Astronomical Observatory; the Partner Group of MPA,
           Nandan Road 80, Shanghai 200030, China}

%%%%%%%%%%%%%%%%%%%%%%%%%%%%%%%%%%%%%%%%%%%%%%%%%%%%%%%%%%%%%%%%%%%%%%%%%%

\date{}

%\pubyear{2004}

\maketitle

\label{firstpage}

%%%%%%%%%%%%%%%%%%%%%%%%%%%%%%%%%%%%%%%%%%%%%%%%%%%%%%%%%%%%%%%%%%%%%%%%%%

\begin{abstract}
  We use the  halo occupation model to calibrate  galaxy group finders
  in  magnitude-limited  redshift surveys.   Since,  according to  the
  current   scenario  of  structure   formation,  galaxy   groups  are
  associated  with  cold  dark  matter  halos, we  make  use  of  the
  properties of the halo population in the design of our group finder.
  The method  starts with an  assumed mass-to-light ratio to  assign a
  tentative mass to each group. This mass is used to estimate the size
  and velocity dispersion of the underlying halo that hosts the group,
  which in  turn is  used to determine  group membership  (in redshift
  space). This procedure is repeated until no further changes occur in
  group memberships. We  find that the final groups  selected this way
  are  insensitive to the  mass-to-light ratio  assumed.  We  use mock
  catalogues,  constructed  using the conditional luminosity  function
  (CLF),  to test the  performance of  our group  finder in  terms of
  completeness of  true members and contamination  by interlopers. Our
  group   finder   is    more   successful   than   the   conventional
  Friends-of-Friends group finder in assigning galaxies in common dark
  matter halos to  a single group.  We apply our  group finder to the
  2-degree  Field Galaxy  Redshift  Survey and  compare the  resulting
  group properties with  model predictions based on the  CLF.  For the
  $\Lambda$CDM  `concordance' cosmology  we find  a  clear discrepancy
  between the model and data in  the sense that the model predicts too
  many rich  groups. In  order to match  the observational  results we
  have to either increase the mass-to-light ratios of rich clusters to
  a level significantly higher than current observational estimates, or
  to assume $\sigma_8 \simeq 0.7$, compared to the `concordance' value
  of $0.9$.
\end{abstract}

%%%%%%%%%%%%%%%%%%%%%%%%%%%%%%%%%%%%%%%%%%%%%%%%%%%%%%%%%%%%%%%%%%%%%%%%%%

\begin{keywords}
dark matter  - large-scale structure of the universe - galaxies:
halos - methods: statistical
\end{keywords}

%%%%%%%%%%%%%%%%%%%%%%%%%%%%%%%%%%%%%%%%%%%%%%%%%%%%%%%%%%%%%%%%%%%%%%%%%%

\section{Introduction}
   
It is common practice to apply a group finder to large galaxy redshift
surveys in order to  assign galaxies to groups\footnote{In this paper,
  we  refer to  a system  of  galaxies as  a group  regardless of  its
  richness, including  isolated galaxies  (i.e., groups with  a single
  member).}.   The  clustering  properties  of galaxies  can  then  be
studied  by analysing  the spatial  clustering of  the groups  and the
distribution functions  of the groups  with respect to  their internal
properties, such as luminosity,  velocity dispersion, mass, shape, and
galaxy population.  Such analyses  have been carried out using various
galaxy redshift surveys, most noticeably  the CfA redshift survey (e.g. 
Geller \& Huchra 1983), the Las Campanas Redshift Survey (e.g.  Tucker
et al.   2000), the 2-degree  Field Galaxy Redshift  Survey (hereafter
2dFGRS; Merch\'an  \& Zandivarez 2002;  Eke et al. 2004a,  2004b), and
the Sloan Digital Sky Survey  (hereafter SDSS; Bahcall \etal 2003; Lee
\etal 2004).

One of  the problems in  such analyses is  that the properties  of the
groups depend on  the group finder used to identify  groups, and it is
in general quite hard to  judge whether the groups identified are true
associations of galaxies in space.  There are several reasons for such
complication. First  of all, galaxies are  observed in redshift-space,
not in the  real space, and so every group finder  has to contend with
the  redshift  distortion  in  the  apparent  clustering  patterns  of
galaxies.    Secondly,   redshift   surveys   are   usually   apparent
magnitude-limited,  and so any  criteria for  clustering based  on the
distances between galaxies  has to take into account  the variation of
the mean inter-galaxy separation  with distance. Thirdly, since only a
finite number of  galaxies can be observed for  each group, shot noise
can also affect the  accuracy of membership assignment, especially for
small groups that contain only a small number of observable galaxies.

One  may argue  that this  is less  a real  problem than  a  matter of
definition  for galaxy groups.  However, since  the ultermate  goal of
such analyses  is to compare with  theory, the definition  must have a
sound physical basis, in order to make a meaningful comparison between
observation and  theory. From the point  of view of  current theory of
galaxy  formation,  the most  natural  reference  for defining  galaxy
groups is  dark matter halos.  These are quasi-equilibrium  systems of
dark  matter   particles,  formed  through   non-linear  gravitational
collapse.  In  the  standard  $\Lambda$CDM model  favored  by  current
observations, most mass at any  given time is bound within dark halos;
galaxies and other luminous objects are assumed to form by cooling and
condensation of the baryons within  halos.  Thus,
it is extremely useful to have  a group finder that can group galaxies
according to  their common dark halos.  Unfortunately,  dark halos are
not directly  observable, and so it  is not possible  to define groups
directly from dark  matter distribution. On the other  hand, since the
current CDM  scenario is  very successful in  explaining a  very large
range of observational data (e.g. Spergel et al. 2003), we may use the
properties  of dark  halos in  current CDM  models as  a guide  in our
design of a group finder. Much  progress has been made in recent years
in  understanding the  relationship  between CDM  halos and  galaxies,
based on  numerical simulations (e.g., Katz, Weinberg \& Hernquist
1996;  Pearce \etal 2000) and semi-analytic models (e.g.,Somerville \& 
Primack 1999; Kauffmann \etal 1999; Cole \etal 2000). 
However, our  understanding for the details about  galaxy formation in
CDM  halos is still quite poor,  and so calibrations  of group finders
have  not been  carried  out  in a  reliable,  model-independent way.  
Furthermore,  to produce  a mock  catalogue that  can not  only  cover a
volume as large as current large redshift surveys of galaxies, such as
2dFGRS and SDSS, but also  have the ability to resolve faint galaxies,
is  not  trivial for  both  numerical  simulation and  semi-analytical
modelling.   What one  really needs  is  an empirical  model that  can
correctly partition the galaxy population into dark halos but does not
depend on uncertain details about galaxy formation in dark halos.

The halo occupation model recently  developed has this spirit. In this
model,  one simply  specifies halo  occupation numbers,  $\langle N(M)
\rangle$, which describe how many galaxies on average occupy a halo of
mass $M$.   Many recent investigations have used  such halo occupation
models  to study  various aspects  of galaxy  clustering (Jing,  Mo \&
B\"orner 1998;  Peacock \& Smith 2000; Seljak  2000; Scoccimarro \etal
2001; White  2001; Jing, B\"orner  \& Suto 2002; Bullock,  Wechsler \&
Somerville 2002;  Berlind \& Weinberg 2002; Scranton  2002; Kang \etal
2002; Marinoni \& Hudson 2002;  Zheng \etal 2002; Magliocchetti \& 
Porciani 2003; Kochanek \etal 2003; Yan, Madgwick\& White 2003; Yan, 
White \& Coil 2004). In two recent papers, Yang, Mo  \& van den Bosch
(2003;  hereafter  Paper~I) and  van  den  Bosch,  Yang \&  Mo  (2003;
hereafter Paper~II) have taken  this halo occupation approach one step
further  by  considering  the  occupation  as  a  function  of  galaxy
luminosity  and  type.   They  introduced the  conditional  luminosity
function (hereafter CLF)  $\Phi(L \vert M) {\rm d}L$,  which gives the
number of galaxies  with luminosities in the range  $L \pm {\rm d}L/2$
that reside in halos of mass  $M$.  The advantage of this CLF over the
halo occupation function $\langle N(M)  \rangle$ is that it allows one
to address the  clustering properties of galaxies {\it  as function of
  luminosity}. In particular, such model  can be used to pupulate dark
matter  halos  in high-resolution  $N$-body  simulations to  construct
realistic  mock galaxy  redshift surveys  that automatically  have the
correct  galaxy  abundances and  correlation  lengths  as function  of
galaxy luminosity and  type. As shown in Yang et  al. (2004), the mock
galaxy  catalogues constructed  in  this  way can  recover  many of  the
properties of  galaxy clustering  in redshift space.  Further analyses
with  such mock catalogues  show that  the model  is also  successful in
matching  the environmental dependence  of galaxy  luminosity function
(Mo et al. 2004), the kinematics of satellite galaxies around galaxies
(van den Bosch et al. 2004ab) and the three-point correlation function
(Wang et al. 2004).

The aim of this paper is two-fold. First, we intend to develop a group
finder that is  based on the consideration given  above, and can group
galaxies according to their common  halos.  We test the reliability of
the method with the extensive  use of various mock samples constructed
from the CLF model. Secondly, we apply the group finder to the 2dFGRS,
and  compare the  properties  of  the groups  so  obtained with  model
predictions.   
The outline  of the paper  is as follows.  In  Section~\ref{sec_GF} we
describe   our   halo-based   group   finder,   which   we   test   in
Section~\ref{sec_testing}  with  the extensive  use  of detailed  mock
samples constructed  from the  CLF model. In  Section~\ref{sec_2dF} we
apply our  group finder to the  2dFGRS, and compare  the properties of
the  groups thus obtained  with model  predictions.  We  summarise our
results in  Section~\ref{sec_conclusion}. 

Unless stated  otherwise, we
consider   a   flat   $\Lambda$CDM  cosmology   with   $\Omega_m=0.3$,
$\Omega_{\Lambda}=0.7$  and  $h=H_0/(100  \kmsmpc)  =  0.7$  and  with
initial  density  fluctuations described  by  a scale-invariant  power
spectrum   with  normalisation  $\sigma_8=0.9$.    These  cosmological
parameters are  in good agreement  with a wide range  of observations,
including the  recent WMAP results  (Spergel \etal 2003), and  in what
follows we refer to it as the ``concordance'' cosmology.

\section{The Group finder}
\label{sec_GF}

Our aim here  is to develop a group finder that  assigns galaxies in a
common halo to  a single group. The properties  of the halo population
in the standard $\Lambda$CDM model are well understood, largely due to
a  combination of  $N$-body simulations  and analytical  models.  Dark
matter  halos  are  defined  as  virialized structures  with  a  mean
over-density of  about 180.  Their density profiles  are well described
by the so-called NFW profile (Navarro, Frenk \& White 1997):
\begin{equation}
\label{NFW}
\rho(r) = \frac{\bar{\delta}\bar{\rho}}{(r/r_{\rm s})(1+r/r_{\rm  s})^{2}},
\end{equation}
where $r_s$  is a characteristic  radius, $\bar{\rho}$ is  the average
density  of  the  Universe,  and  $\bar{\delta}$  is  a  dimensionless
amplitude which  can be expressed  in terms of the  halo concentration
parameter $c=r_{180}/r_s$ as
\begin{equation}
\label{overdensity}
\bar{\delta} = {180 \over 3} \, {c^{3} \over {\rm ln}(1+c) - c/(1+c)}.
\end{equation}
Here  $r_{180}$ is the  radius within  which the  halo has  an average
over-density of  $180$. Numerical simulations show that the halo 
concentration depends on halo mass, we use the relation given by
Bullock \etal (2001) and properly rescaled to our definition.

Both observations and numerical simulations suggest that the brightest
galaxy in each  dark matter halo resides at rest  at the centre, while
the  number density  distribution of  the fainter,  satellite galaxies
matches that of the dark  matter particles (e.g., Carlberg \etal 1997;
van der  Marel \etal 2000; Berlind  \etal 2003; Lin,  Mohr \& Stanford
2004; Rines  \etal 2004; Diemand \etal  2004).  Furthermore, although
velocity bias may exist in the sense that galaxies move with different
velocities as  the dark  matter particles at  the same  location, such
bias is  found to  be not large  (Berlind \etal 2003,  Yoshikawa \etal
2003).   These  findings  suggest  that  it may  be  possible  to  use
information regarding the spatial distribution and velocity dispersion
of  galaxy group  members to  estimate halo  masses, and  vice  versa. 
Motivated  by these  considerations,  we design  a  group finder  that
consists of the following steps:

{\bf Step 1:} We combine two different methods to identify the centres
of potential  groups. First we use  the traditional Friends-Of-Friends
(FOF) algorithm to assign galaxies  into groups.  Since we are working
in redshift space, we separately define linking lengths along the line
of sight ($\ell_z$) and in the transverse direction ($\ell_p$).  Since
the  purpose  here is  only  to identify  the  group  centres, we  use
relatively small linking lengths: $\ell_z=0.3$ and $\ell_p=0.05$, both
in units of the mean separation of galaxies. Note that for an apparent
magnitude limited survey the mean separation of galaxies is a function
of redshift, which  we take into account.  The  geometrical centres of
all  FOF  groups  thus  identified  with  more  than  2  galaxies  are
considered as centres of potential groups. Next, from all galaxies not
yet linked together by these  FOF groups, we select bright, relatively
isolated  galaxies  which  we  also  associate  with  the  centres  of
potential  groups. Following McKay \etal  (2002), Prada  \etal (2003),
Brainerd \& Specian (2003) and van den Bosch \etal (2004a), we identify
a galaxy  as `central', and thus  as the centre of  a potential group,
when it is  the brightest galaxy in a cylinder  of radius $1\mpch$ and
velocity depth $\pm 500\kms$.

{\bf Step 2:} We estimate the luminosity of a selected potential group
using
\begin{equation}
L_{\rm group} = \sum_i \frac{L_i}{f_c(L_i)} 
\end{equation}
where $L_i$ is  the luminosity of each galaxy in  the group, and $f_c$
is  the luminosity-dependent  incompleteness of  the survey  (which is
relevant only when  the group finder is applied to  real data, such as
the 2dFGRS).  The total luminosity of the group is approximated by
\begin{equation}\label{eq:L_total}
L_{\rm total} = L_{\rm group} \frac{\int_0^{\infty} L\phi(L)dL}
{\int_{L_{\rm lim}}^{\infty} L\phi(L)dL}\,,
\end{equation}
where $L_{\rm lim}$ is the minimum  luminosity of a galaxy that can be
observed at  the redshift  of the group,  and $\phi(L)$ is  the galaxy
luminosity  function   
\begin{equation}
\phi(L)dL = \phi_* \left (\frac{L}{L_*}\right )^{\alpha} \exp 
\left ( -\frac{L}{L_*}\right )
\frac{dL}{L_*}\,.
\end{equation}
Throughout  this  paper, galaxy luminosities are defined
in the photometric $b_J$ band. For the luminosity function,
we take the  parameters from  Norberg \etal  (2002):  
($\phi_*, M_*, \alpha$) = ($0.0161$, $-19.66$, $-1.21$).

{\bf  Step  3:}  From  $L_{\rm  total}$  and a  model  for  the  group
mass-to-light ratio  (see below), we  compute an estimate of  the halo
mass associated with the group in consideration. From this estimate we
also compute the halo radius  $r_{\rm 180}$, the virial radius $r_{\rm
  vir}$\footnote{The virial radius is  defined as the radius inside of
  which the  average density is $\Delta_{\rm vir}$  times the critical
  density, with  $\Delta_{\rm vir}$ given by Bryan  \& Norman (1998)},
and the  virial velocity $V_{\rm  vir} = (G  M / r_{\rm  vir})^{1/2}$. 
The line-of-sight velocity dispersion  of the galaxies within the dark
matter halo is assumed to be $\sigma = V_{\rm vir}/\sqrt{2}$.
  
{\bf Step 4:} Once we have a group centre, and a tentative estimate of
the group size,  mass, and velocity dispersion, we  can assign galaxies
to this  group according to the  properties of the  associated halos. 
If  we assume that  the phase-space  distribution of  galaxies follows
that  of the  dark matter  particles, the  number density  contrast of
galaxies in redshift space around the group centre ($=$ centre of dark
matter halo) at redshift $z_{\rm group}$ can be written as
\begin{equation}
P_M(R,\Delta z) = {H_0\over c}
{\Sigma(R)\over {\bar \rho}} p(\Delta z) \,,
\end{equation}
Here $\Delta z  = z - z_{\rm group}$ and  $\Sigma(R)$ is the projected
surface density of a (spherical) NFW halo:
\begin{equation}
\Sigma(R)= 2~r_s~\bar{\delta}~{\bar \rho}~{f(R/r_c)}\,,
\end{equation}
with 
\begin{equation}\label{eq:fx}
f(x) = \left\{
\begin{array}{lll}
\frac{1}{x^{2}-1}\left(1-\frac{{\ln
{\frac{1+\sqrt{1-x^2}}{x}}}}{\sqrt{1-x^{2}}}\right)
 & \mbox{if $x<1$} \\
\frac{1}{3}
 & \mbox{if $x=1$} \\
\frac{1}{x^{2}-1}\left(1-\frac{{\rm
atan}\sqrt{x^2-1}}{\sqrt{x^{2}-1}}\right)
 & \mbox{if $x>1$}
\end{array} \right.\,.
\end{equation}
The  function $p(\Delta  z){\rm d}\Delta  z $  describes  the redshift
distribution of galaxies within the halo for which we adopt a Gaussian
form
\begin{equation}
p(\Delta z)= \frac{c}{\sqrt{2\pi}\sigma (1+z_{\rm group})} 
\exp \left [ \frac {-(c\Delta z)^2}
{2\sigma^2(1+z_{\rm group})^2}\right ] \,,
\end{equation}
where $\sigma$ is the rest-frame velocity dispersion.

Thus  defined,  $P_M(R,\Delta  z)$  is the  three-dimensional  density
contrast  in redshift  space.  In  order  to decide  whether a  galaxy
should be assigned  to a particular group we  proceed as follows.  For
each galaxy  we loop  over all groups,  and compute  the corresponding
distance $(R,\Delta z)$ between galaxy  and group centre.  Here $R$ is
the projected distance at the  redshift of the group. If $P_M(R,\Delta
z)  \ge B$,  with $B$  an appropriately  chosen background  level, the
galaxy is assigned  to the group. If a galaxy can  be assigned to more
than  one  group,  it  is   only  assigned  to  the  group  for  which
$P_M(R,\Delta z)$ has  the highest value.  Finally, if  all members of
two  groups  can be  assigned  to one  group  according  to the  above
criterion,  the  two groups  are  merged  into  a single  group.   The
background level $B$ defines  a threshold density contrast in redshift
space  whose  value  we  calibrate  using mock  galaxy  catalogues  (see
Section~\ref{sec:test2}).

{\bf Step 5:}  Using the group members thus  selected we recompute the
group-centre  and go  back  to Step  2,  iterating until  there is  no
further change in  the memberships of groups.  Note  that, unlike with
the traditional  FOF method, this group finder  also identifies groups
with only one member.

The basic idea of this group  finder is similar to that of the matched
filter algorithms  developed by Postman \etal (1996)  (see also Kepner
\etal 1999;  White \&  Kochanek 2002; Kim  \etal 2002;  Kochanek \etal
2003),  although we  also make  use of  the galaxy  kinematics  in our
analyses. The crucial ingredient of  our method is to somehow obtain a
reliable estimate  of the halo  mass associated with the  galaxy group
from observable properties of  its selected member galaxies.  Ideally,
this  mass estimate  should  be model-independent.   For example,  one
might hope  to estimate the mass  from the velocity  dispersion of the
selected   members,  using   the  fact   that  $M\propto   \sigma^3$.  
Unfortunately, we  found that  this is not  a reliable  method, simply
because  the estimate of  $\sigma$ is  too noisy  when only  few group
members are  available. Therefore, we decided  to use a  model for the
mass-to-light  ratio  to  estimate  the  group  mass  from  the  total
luminosity of the selected  galaxies.  Since total group luminosity is
dominated by  the few brightest  galaxies, the estimated mass  is much
less sensitive to  the absence of faint group  members.  A downside of
this  method,   however,  is  that   it  requires  a  model   for  the
mass-to-light ratios.   Fortunately, we found that  the memberships of
the selected groups  are remarkably insensitive to the  adopted model. 
The reason for this is that, even  if the estimated mass is wrong by a
factor of 10, the implied  radius and velocity dispersion, used in the
membership determination, change only by a factor of $2.15$

In what follows we use  the average mass-to-light ratios obtained by
van den Bosch, Yang \& Mo (2003) from the CLF formalism
\begin{equation}
\label{eq:ml}
\frac{M}{L} = \left\{
\begin{array}{lll}
\frac{1}{2} \left ( \frac{M}{L} \right )_0 
 \left [ \left ( \frac{M}{M_1}\right )^{-\gamma_1} +
  \left ( \frac{M}{M_1}\right )^{\gamma_2} \right ] 
 & \mbox{if $M_{14} \leq 1 $} \\
{(M/L)_{cl}} & \mbox{if  $M_{14} > 1$}
\end{array} \right.
\end{equation}
where we  set the model parameters  to those of their  model~D: $M_1 =
10^{10.94}    h^{-1}    \Msun$,   $\gamma_1=2.02$,    $\gamma_2=0.30$,
$(M/L)_0=124 h \MLsun$, and $(M/L)_{cl}=500 h \MLsun$. $M_{14}$ is the
halo mass in units of $10^{14} h^{-1} \Msun$. As we demonstrate in the
next section, even if we  adopt a constant mass-to-light ratio in Step
3 of  $M/L =  400 h \MLsun$  for all  halos, the selected  groups are
virtually  identical  to those  selected  when  using these  CLF-based
mass-to-light ratios.

\section{Testing the group finder}
\label{sec_testing}

\begin{figure*}
\centerline{\psfig{figure=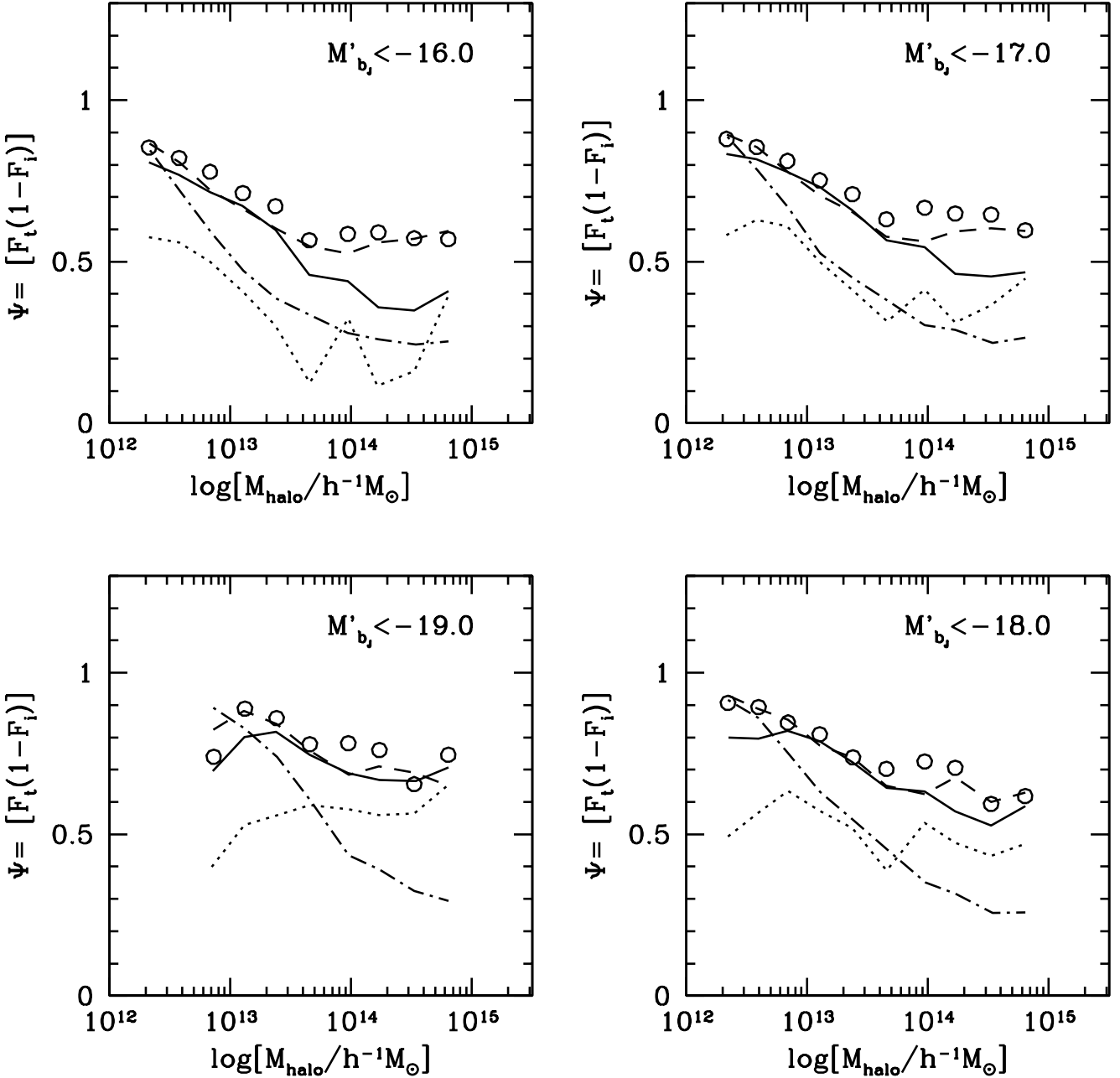,width=\hdsize}}
\caption{The value of $\Psi \equiv F_{\rm t} \times (1-F_{\rm i})$ 
  (where $F_{\rm t}$ is the completeness factor and $F_{\rm i}$ is the
  contamination  factor) as  a  function of  halo  mass for  different
  choices of  the background level  $B$.  The circles show  the result
  for our  fiducial group  finder with $B=10$.   The solid  and dashed
  lines correspond  to $B=5$ and $20$, respectively,  while the dotted
  and dot-dashed lines outline two extreme cases with $B=0$ and $100$,
  respectively.   The   four  different  panels   show  results  using
  different luminosity cuts: only  galaxies brighter than the absolute
  magnitude  limit indicated  (where  $M'_{b_J}=M_{b_J}-5\log h$)  are
  used to identify groups and to define $F_{\rm t}$ and $F_{\rm i}$.}
\label{fig:ftfi}
\end{figure*}
\begin{figure*}
\centerline{\psfig{figure=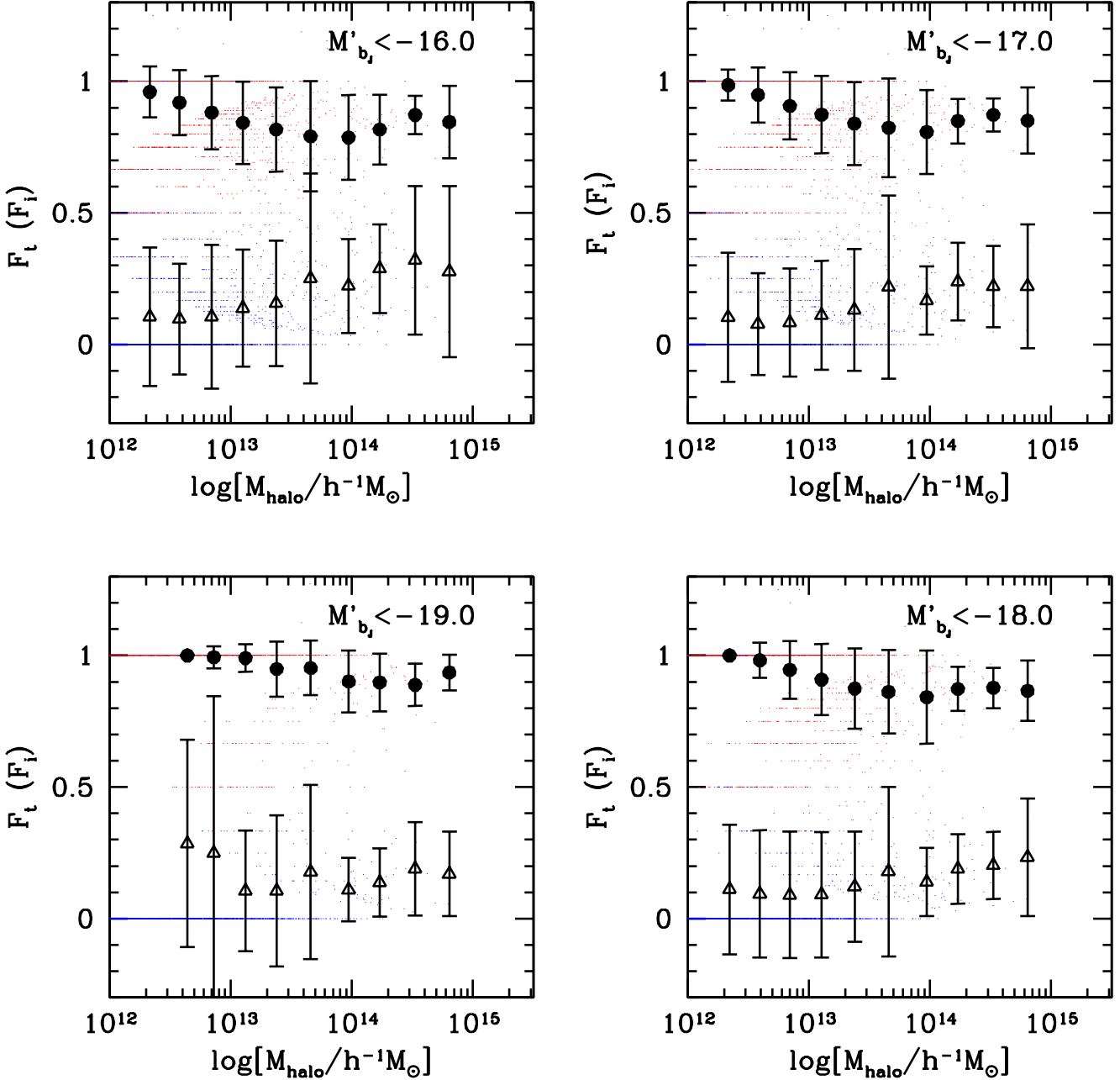,width=\hdsize}}
\caption{Red and blue points indicate the completeness and
  contamination,  respectively, of  individual groups  as  function of
  halo mass. Solid dots  and open triangles indicate the corresponding
  averages  with the  errorbars  indicating the  1-$\sigma$ variance.  
  Results  correspond  to  our  fiducial  group finder  with  $B=10$.  
  Different panels  correspond to different  absolute magnitude limits
  as in Fig.~\ref{fig:ftfi}.}
\label{fig:complete}
\end{figure*}
\begin{figure*}
\centerline{\psfig{figure=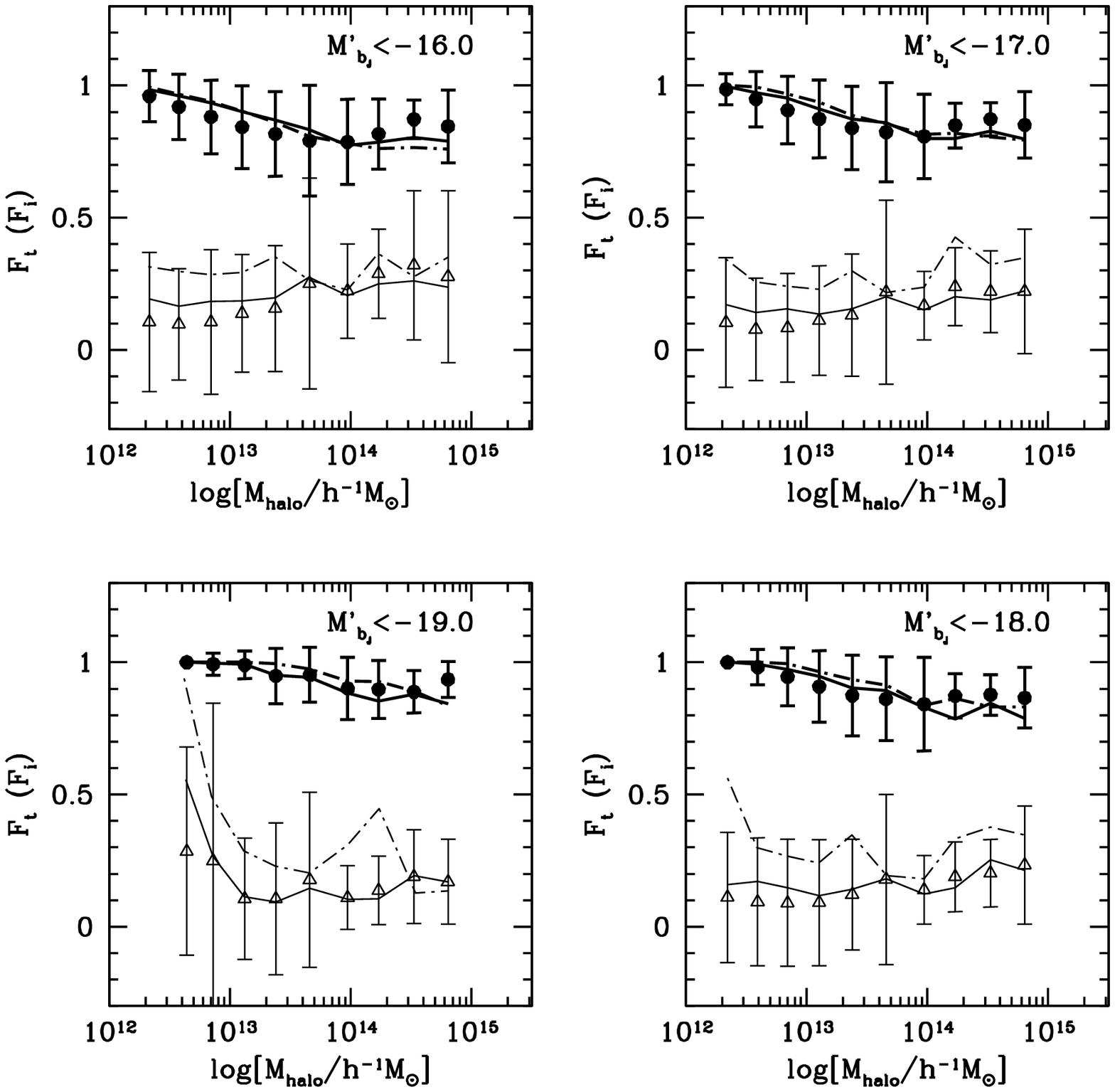,width=\hdsize}}
\caption{Same as in Fig.~\ref{fig:complete}, but here we compare 
  different group finder methods.   The thick (thin) lines and symbols
  show the completeness (contamination) of the groups. The symbols 
  with errorbars are  the same  results as in Fig.~\ref{fig:complete}.   
  The solid lines show  the  results  in which  we  assume a  constant
  mass-to-light   ratio   $M/L=400   (h  M_{\odot}/L_{\odot})$.    The
  dot-dashed lines show  the results where we use  the traditional FOF
  group finder  method with the same  linking lengths as  in Eke \etal
  (2004a).  
}
\label{fig:fof}
\end{figure*}

\subsection{Constructing Mock Samples}
\label{sec:construct}

We construct mock  galaxy samples by populating dark  matter halos in
numerical simulations  with galaxies of  different luminosities, using
the  conditional luminosity function  formalism developed  in Papers~I
and~II. In what  follows we focus on CLF  model~D defined in paper~II,
which  is valid  for the  concordance cosmology  considered  here, and
which  yields excellent  fits to  the  observed LFs  and the  observed
correlation lengths as function of both luminosity and type.  The same
CLF has also  been used in Yang \etal (2004)  to construct large, mock
galaxy  redshift surveys,  which they  used to  compare  various large
scale structure statistics with those obtained from the 2dFGRS.

In this  paper we use the same  mock samples to test  our group finder
and to construct detailed  mock galaxy redshift surveys for comparison
with the 2dFGRS.  Here we briefly summarise the ingredients, but refer
the reader  to Yang  \etal (2004) for  details.  The mock  samples are
constructed using  a set of $N$-body  simulations carried out  by Y.P. 
Jing and  Y.  Suto (see Jing 2002;  Jing \& Suto 2002)  on the VPP5000
Fujitsu  supercomputer  of the  National  Astronomical Observatory  of
Japan using a vectorized-parallel P$^3$M  code.  The set consists of a
total of  six simulations, each  of which uses $N=512^3$  particles to
evolve the  distribution of  dark matter from  an initial  redshift of
$z=72$ down  to $z=0$ in a $\Lambda$CDM  `concordance' cosmology.  All
simulations consider  boxes with periodic boundary  conditions; in two
cases $L_{\rm  box}=100 h^{-1} \Mpc$ while the  other four simulations
all have  $L_{\rm box}=300  h^{-1} \Mpc$.  Different  simulations with
the same box size are completely independent realizations and are used
to estimate uncertainties due to cosmic variance.  The particle masses
are $6.2  \times 10^8 \msunh$  and $1.7\times 10^{10} \msunh$  for the
small and large box simulations, respectively.  Dark matter halos are
identified  using the  FOF algorithm  with a  linking length  of $0.2$
times  the  mean   inter-particle  separation.   For  each  individual
simulation we  construct a  catalogue of halos  with $10$  particles or
more, for  which we  store the  mass, the position  of the  most bound
particle, and the  halo's mean velocity and velocity dispersion. Halos
that are unbound are removed from  the sample. In Yang \etal (2004) we
have shown  that the  resulting halo mass  functions are  in excellent
agreement with the analytical halo mass    function given by Sheth  \& 
Tormen (1999) and    Sheth, Mo \& Tormen  (2001).    The  mock  galaxy
distributions are  constructed by  populating the dark matter halos in
these  $N$-body simulations with  galaxies according  to the  CLF. The
brightest galaxy in each halo is  located at rest at the centre of the
halo, while  the other galaxies  follow a number  density distribution
that is identical to the mass distribution of the dark matter and with
an isotropic velocity dispersion that is identical to that of the dark
matter.   Finally, when  using these  mock  samples to  create a  mock
2dFGRS (hereafter MGRS), we stack simulations with different box sizes
and resolutions in order to cover a large redshift range (to match the
depth  of the  2dFGRS)  and to  properly  sample the  faint galaxies.  
Observational  errors and  selection effects  in the  2dFGRS,  such as
position-dependent completeness and magnitude-limit variations are all
taken into account.  From our set of six simulation boxes we construct
eight independent MGRSs, which we use to estimate the impact of cosmic
variance.
\begin{figure*}
\centerline{\psfig{figure=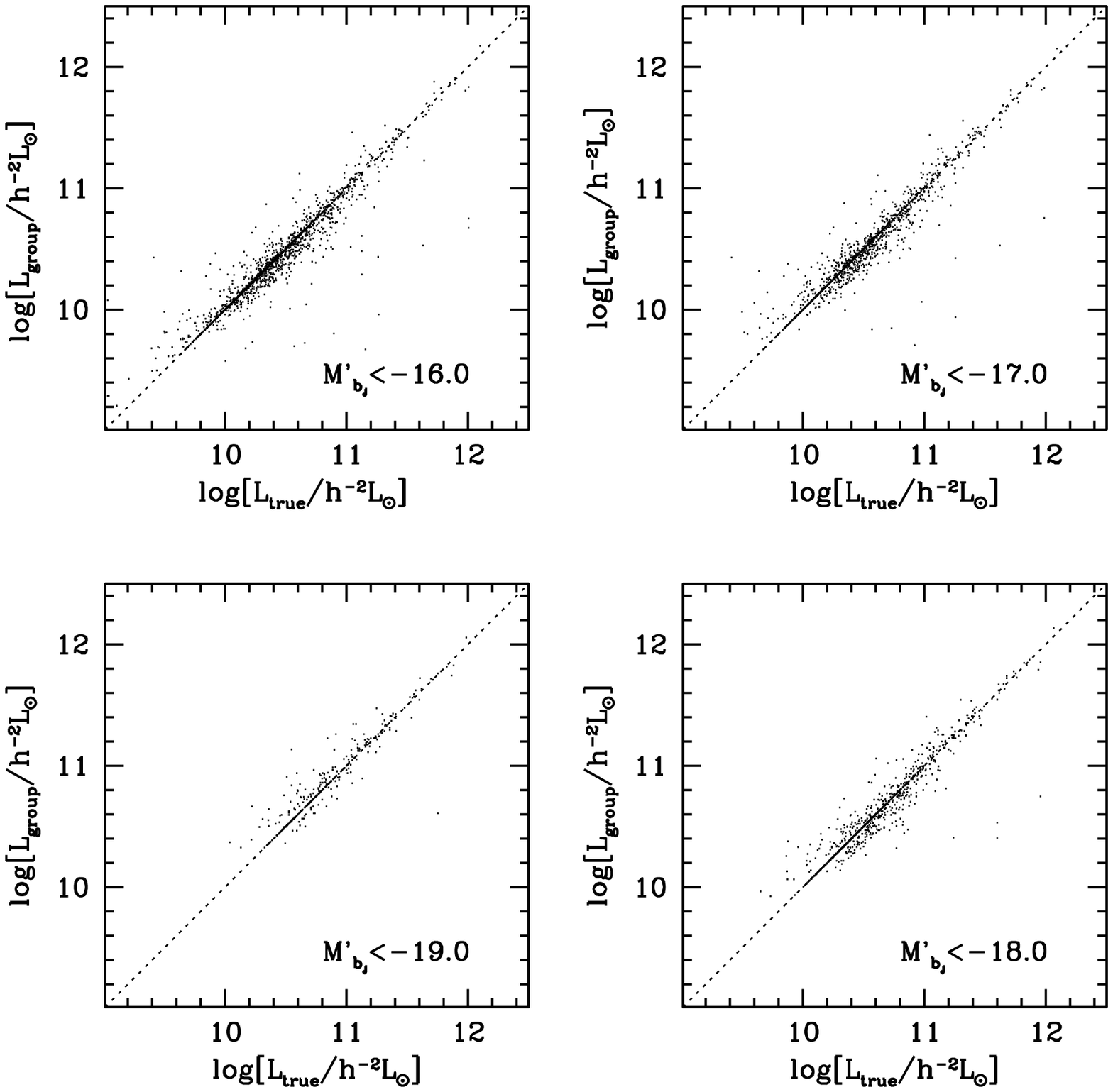,width=\hdsize}}
\caption{Comparison of the group luminosity $L_{\rm group}$
  (the total luminosity of all identified group members) with the {\it
    true} group luminosity $L_{\rm true}$ (the total luminosity of all
  true  group members).  Only  results  for groups  with  more than  2
  members are shown. The four different panels correspond to different
  absolute      magnitude     cuts      (as      indicated),     where
  $M'_{b_J}=M_{b_J}-5\log h$.}
\label{fig:L_L}
\end{figure*}
\begin{figure*}
\centerline{\psfig{figure=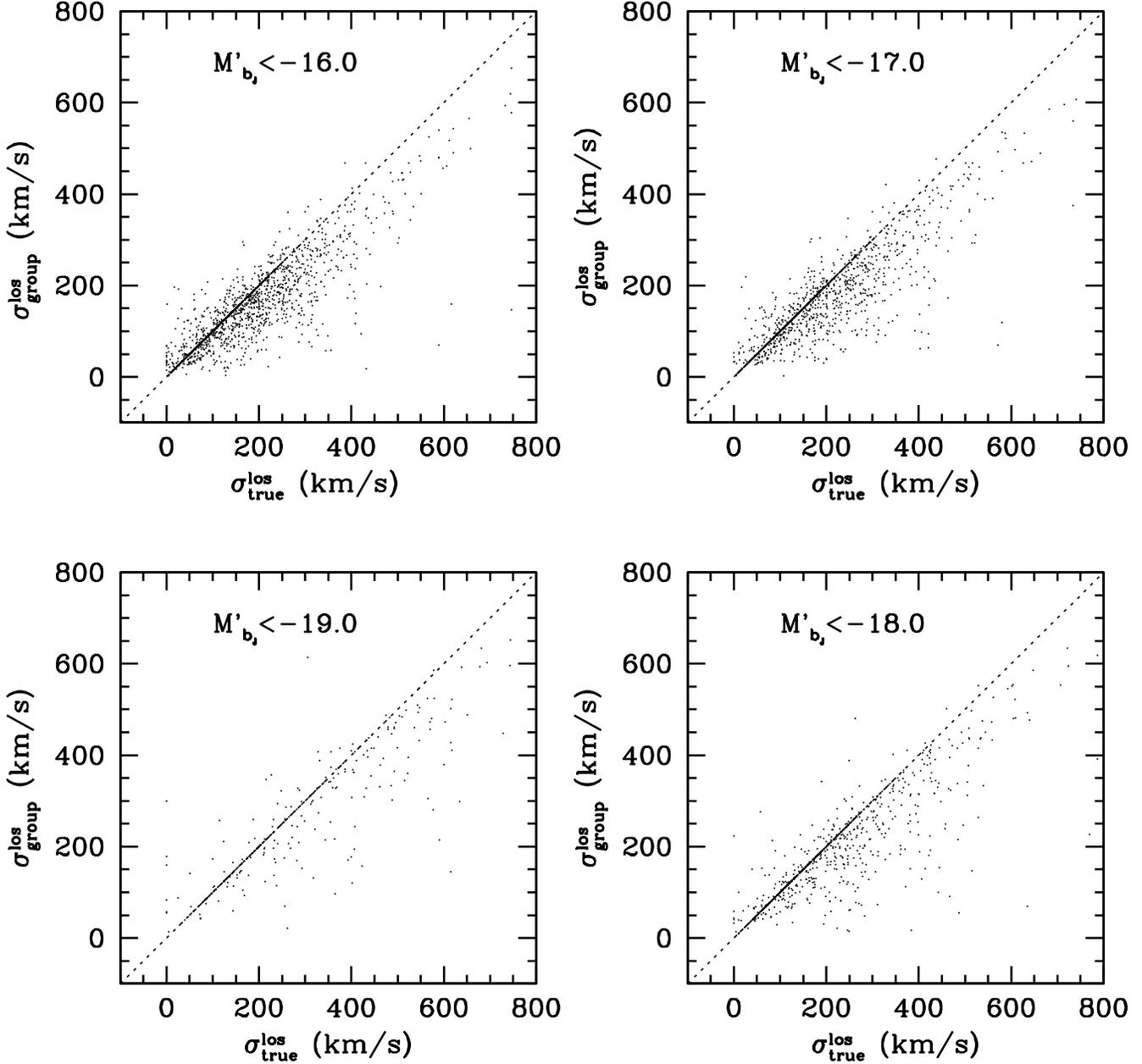,width=\hdsize}}
\caption{Same as Fig.~\ref{fig:L_L}, except that here we compare the 
  line-of-sight velocity  dispersion of identified  group members with
  the line-of-sight velocity dispersion  of true members. All velocity
  dispersions are  estimated using  the gapper estimator  described in
  the text.}
\label{fig:v_v}
\end{figure*}
\begin{figure*}
\centerline{\psfig{figure=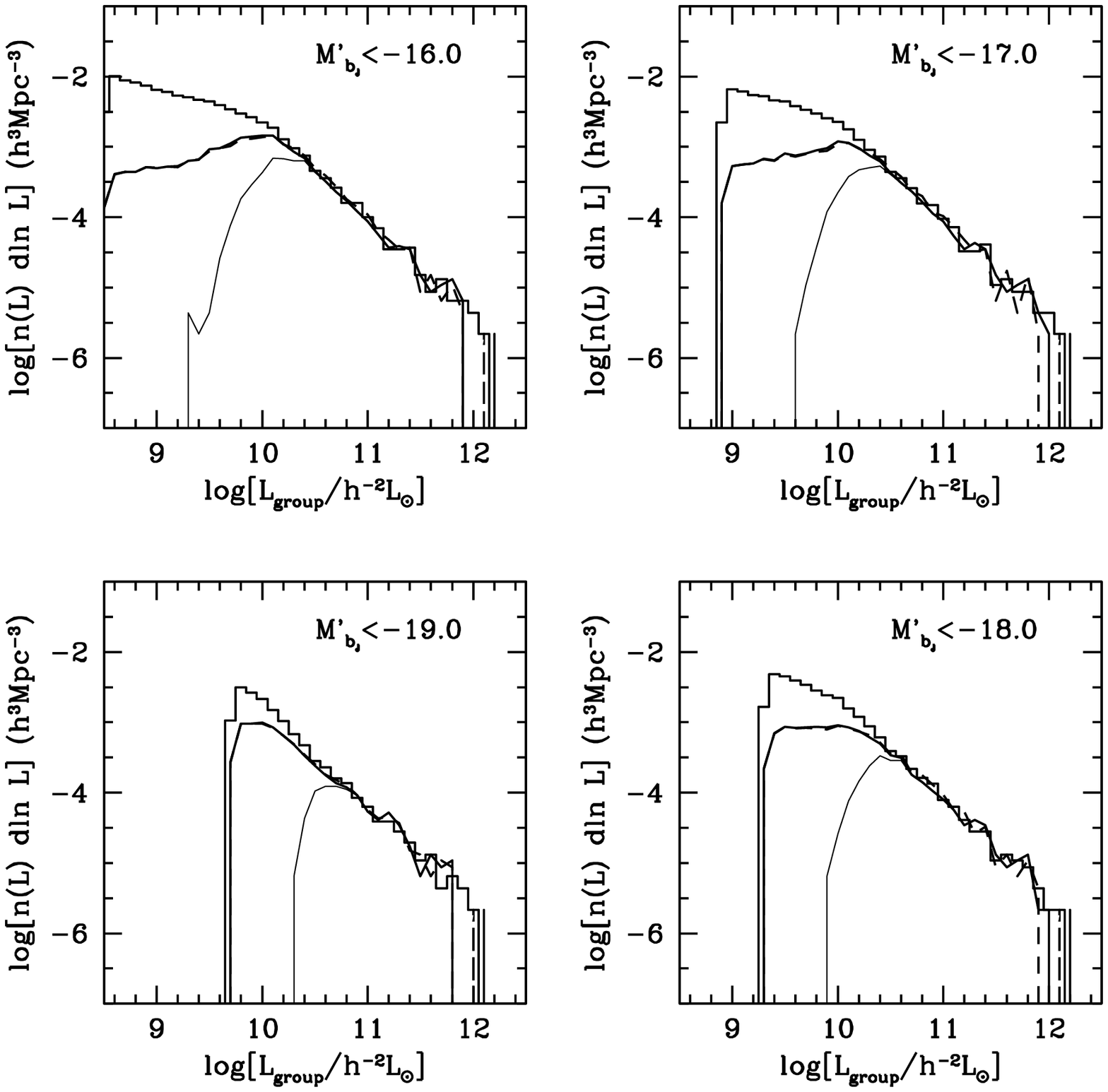,width=\hdsize}}
\caption{Group luminosity functions. Histograms indicate the 
  true luminosity functions, in which  the luminosity of each group is
  the sum of the luminosities  of all true group members brighter than
  the  luminosity  cut  indicated  in  each  panel.   The  solid  lines
  correspond to the luminosity functions for all the groups identified
  with our fiducial group finder,  while the thin solid lines indicate
  the  corresponding  luminosity functions  obtained  when only  using
  groups with at least  three identified  members.  The  dashed lines,
  which almost overlap with the thick solid lines,
  indicate the group luminosity functions obtained when using
  a constant mass-to-light ratio of  $M/L=400 (h \MLsun)$ in the group
  finder.}
\label{fig:LF}
\end{figure*}

\subsection{Volume limited samples}
\label{sec:test2}

Before applying our group finder to the MGRSs and the 2dFGRS, we first
test and calibrate  it using a volume limited  sample of mock galaxies
in one of our simulation boxes with $L_{\rm box}=100 h^{-1} \Mpc$.  To
mimic observation, we simply assume  redshifts to be along one side of
the simulation box.

In what follows  a {\it true} group is defined as  the set of galaxies
that reside in  the same halo.  In order to  quantify the group finder
performance we introduce the completeness, $F_{\rm t}$, defined as the
ratio between the number of  true members selected by the group finder
and the  total number  of true group  members, and  the contamination,
$F_{\rm i}$, defined as the  ratio between the number of false members
(interlopers) selected  by the  group finder and  the total  number of
true members.   A perfect group  finder has $F_{\rm t}=1$  and $F_{\rm
  i}=0$.  Therefore,  we optimise  our group finder  by simultaneously
maximising $F_{\rm  t}$ and minimising  $F_{\rm i}$.  Note  that these
two criteria  may sometimes  be in conflict.   For example,  a certain
group  finder  may select  all  true members  but  at  the expense  of
including  many  interlopers.   Other  group finders  may  yield  zero
interlopers but  miss a large  fraction of true members.   Whether one
gives more weight to large $F_{\rm t}$ or small $F_{\rm i}$ depends on
the  question one  wishes to  address.  In  this paper  we  give equal
weight  to both  criteria,  and we  tune  our parameter  in the  group
finder,  i.e.   the  background  level  $B$,  so  that  the  value  of
$\Psi\equiv  F_{\rm   t}\times  (1-F_{\rm  i})$   is  maximised.   The
background  level $B$  defined in  this way  is about  $10$  for $M_*$
halos, and $\Psi$ has a weak dependence on the mass of the system. 

The performance of our group finder is not very sensitive to the exact
value  of  $B$  used. This  is  evident  from  the various  curves  in
Fig.~\ref{fig:ftfi},  which show  that  similar values  of $\Psi$  are
obtained using a  fairly large range of background  levels. To put the
absolute values of  $B$ in perspective we recall that it  is used as a
threshold for the redshift-space  density contrast of groups. Ideally,
$B$ should therefore correspond  roughly to the redshift-space density
contrast at the edge of a halo, i.e.,
\begin{equation}
B \approx {\rho_{\rm red}(r_{180}) \over \bar{\rho}} 
  \approx { \rho(r_{180}) \over \bar{\rho} }
  {(4\pi /3) r_{180}^3 \over \pi r_{180}^2 \sigma /H_0}\, 
\end{equation}
Using that  ${\rho(r_{180}) / \bar{\rho}}\sim  30$ and ${\sigma  / H_0
  r_{180}} \sim 4$, we obtain $B \sim 10$, in excellent agreement with
the value  obtained by maximising $\Psi$.  Therefore,  in what follows
we adopt $B=10$, independent of halo mass.
 
Fig.~\ref{fig:complete}  gives   a  more  detailed   overview  of  the
completeness  and contamination  for  our fiducial  group finder  with
$B=10$. Dots  correspond to individual  groups (halos), while  the big
symbols  with  errorbars  indicate  the  average  $F_{\rm  t}$  (solid
circles) and $F_{\rm i}$ (open triangles). Only galaxies brighter than
a  certain  absolute-magnitude limit  (indicated  in  the upper  right
corner of each panel) are used to identify groups.  The incompleteness
and contamination are  also defined with respect to  all group members
brighter  than  that  limit.    Overall  the  group  finder  is  quite
successful, with an average completeness  of about $90\%$ 
and a contamination of about $20\%$ for groups in
dark matter  halos with masses  spanning the entire range  from $\sim
10^{12}h^{-1}{\rm  M}_\odot$  to  $\sim 10^{15}h^{-1}{\rm  M}_\odot$.  
Note  also  that the  completeness  and  contamination  do not  depend
significantly  on the luminosity  limit of  the tracer  galaxies. This
implies that our group finder can make a fairly uniform identification
of galaxy groups even in  an apparent-magnitude limited sample.  As we
shall see in Section~\ref{sec:test_mag}, this is indeed the case.

In the group finder used thus far we have used the mass-to-light ratio
predicted  by the CLF  (eq.~[\ref{eq:ml}]) to  estimate the  sizes and
velocity dispersions of the groups.   In order to test the sensitivity
of the group finder to this model assumption, we performed a number of
tests  with different mass-to-light  ratios. One  of the  more extreme
examples tested is a model with constant mass-to-light ratio of $M/L =
400 h \MLsun$ independent of halo mass. As shown in Fig.~\ref{fig:fof}
the  resulting   completeness   and   contamination  are  very similar
to  those obtained using our fiducial mass-to-light
ratios  of  eq.~(\ref{eq:ml}). We  obtained  similar  results for  all
mass-to-light  ratios  tested, indicating  that  the completeness  and
contamination levels of our  group finder are extremely insensitive to
the exact mass-to-light ratios assumed.

Fig.~\ref{fig:fof} also shows the  results for groups identified using
the standard friends-of-friends (FOF) algorithm (i.e., only using Step
1 in  Section~\ref{sec_GF}).  We use exactly the  same linking lengths
as in Eke \etal (2004a):
\begin{equation}
\label{eq_link1}
\ell_p=0.13~(\Delta/5)^{0.04}~n^{-1/3}
\end{equation}
in   the   transverse   direction,   and
\begin{equation}
\label{eq_link2}
\ell_z=1.43~(\Delta/5)^{0.16}~n^{-1/3}
\end{equation}
along the line of sight, with  $n$ the number density of galaxies. The
quantity  $\Delta$ is  the  galaxy density  contrast  relative to  the
background  at  the  redshift  considered  in  a  cylinder  of  radius
$1.5\mpch$  and   velocity  depth  $\pm  1650   \kms$.   Although  the
completeness  given by  the  FOF  algorithm is  similar  to our  group
finder,  the contamination  is significantly  higher. In  addition the
contamination in poor groups  increases systematically when only using
brighter galaxies.  Therefore, the FOF  method does not give a uniform
identification in an apparent magnitude limited sample (e.g.  Diaferio
\etal 1999).

To  further test  the  performance  of our  group  finder, we  compare
properties of  the identified  groups with those  of the true  groups. 
Fig.~\ref{fig:L_L}  shows  the   predicted  group  luminosity  $L_{\rm
  group}$ (the sum of the  luminosities of all members assigned by the
group finder) versus the true group luminosity $L_{\rm true}$ (the sum
of the  luminosities of  all true members).  We only show  results for
groups with three  or more selected members.  Each  of the four panels
shows the result where galaxies brighter than a certain absolute 
magnitude limit are used  in selecting the groups and
in  calculating the  group luminosities.   Overall the  correlation is
fairly tight,  indicating that the incompleteness and  the presence of
interlopers does not strongly impact on total group luminosity.

For a system in dynamical  equilibrium, its mass may be estimated from
its velocity dispersion through the virial theorem. Since the velocity
dispersion of  a group  can be measured  from the redshifts  of member
galaxies,  it is interesting  to examine  how the  velocity dispersion
among selected  members compares to that  of the true  members. We use
the gapper estimator  described by Beers, Flynn \&  Gebhardt (1990) to
estimate the group velocity dispersions.  The method involves ordering
the set of recession velocities $\{v_i\}$ of the $N$ group members and
defining gaps as
\begin{equation}
g_i=v_{i+1}-v_i,~~~ i=1,2,...,N-1.
\end{equation}
The rest-frame velocity dispersion is then estimated by
\begin{equation}
\sigma_{\rm gap}=\frac{\sqrt{\pi}}{(1+z_{\rm group})N(N-1)}
\sum_{i=1}^{N-1}w_i g_i\,.
\label{eq:vgap}
\end{equation}
where the weight is defined as $w_i=i(N-i)$.  Since there is a central
galaxy in each group (which is  true in our mock samples, and which we
also  assume to  hold in  the real  Universe), the  estimated velocity
dispersion has  to be  corrected by a  factor of  $\sqrt{N/(N-1)}$. In
addition, both  in the  2dFGRS and  in our MGRSs  there is  a redshift
measurement  error of  $\sigma_{\rm  err}=85\kms$, which  needs to  be
removed in  quadrature.  Thus the  final velocity dispersion  of group
members is given by
\begin{equation}\label{eq_gapper}
\sigma=\sqrt{{\rm max}\left(0,\frac{N\sigma_{\rm gap}^2}{N-1}
-\sigma_{\rm err}^2\right)}\,
\label{eq:sigma}
\end{equation}
(Eke \etal 2004a).   Fig.~\ref{fig:v_v} shows the velocity dispersions
of  groups   obtained  from  the   identified  galaxies,  $\sigma^{\rm
  los}_{\rm  group}$, versus  those  obtained from  the true  members,
$\sigma^{\rm los}_{\rm  true}$. As in  Fig.~\ref{fig:L_L}, only groups
with three  of more members  are shown. Contrary to  the luminosities,
the  scatter here  is quite  large, and  there is  a clear  trend that
$\sigma^{\rm  los}_{\rm group}  < \sigma^{\rm  los}_{\rm  true}$. This
discrepancy is due to the fact that galaxies with the highest peculiar
velocities in  a group are the most  likely to be missed  by the group
finder due to the incompleteness.  We can reduce the
fraction of missed  members by lowering the background  level $B$, but
at  the  expense  of  a   significant  increase  in  the  fraction  of
interlopers.  This indicates that it  is in general quite difficult to
accurately estimate the depth of the gravitational potential well of a
dark matter halo from the velocities of its member galaxies identified
in redshift space.
  
The above  tests are based on  the properties of  selected groups, but
they do not  address whether or not the group finder  can find all the
existing   groups.    To  check   the   completeness   of  the   group
identification, we estimate the group luminosity function, which gives
the  number density  of  groups as  a  function of  group luminosity.  
Fig.~\ref{fig:LF} shows the  group luminosity functions estimated from
selected groups  with at  least 1  (thick lines) or  at least  3 (thin
lines) members. The histogram indicates the luminosity function of all
true  groups (i.e.,  of  all dark  matter  halos), and  is shown  for
comparison.  If we  include all systems selected by  our groups finder
(single  galaxies,  binary  systems,  and  groups  with  more  than  2
members), the group catalogue is roughly complete down to a luminosity
of about $10^{10.5}h^{-2}{\rm L}_\odot$ (corresponding to a typical halo
mass of $\sim 10^{12.5}h^{-1}{\rm M}_\odot$). If, on the other hand, we
exclude single galaxies and binary systems, the sample is only more or
less complete down to a group luminosity of about $10^{10.8}h^{-2}{\rm
  L}_\odot$.    Finally,  we   tested  the   impact  of   the  assumed
mass-to-light ratio on the completeness of the group finder by using a
constant  $M/L=400  h M_{\odot}/L_{\odot}$.   This  leads  to a  group
luminosity function that differs  only very slightly from the fiducial
case (dashed lines Fig.~\ref{fig:LF}),  indicating once again that our
group  finder   is  insensitive  to  the   assumptions  regarding  the
mass-to-light ratios.

\subsection{Flux limited samples}
\label{sec:test_mag}

\begin{figure*}
\centerline{\psfig{figure=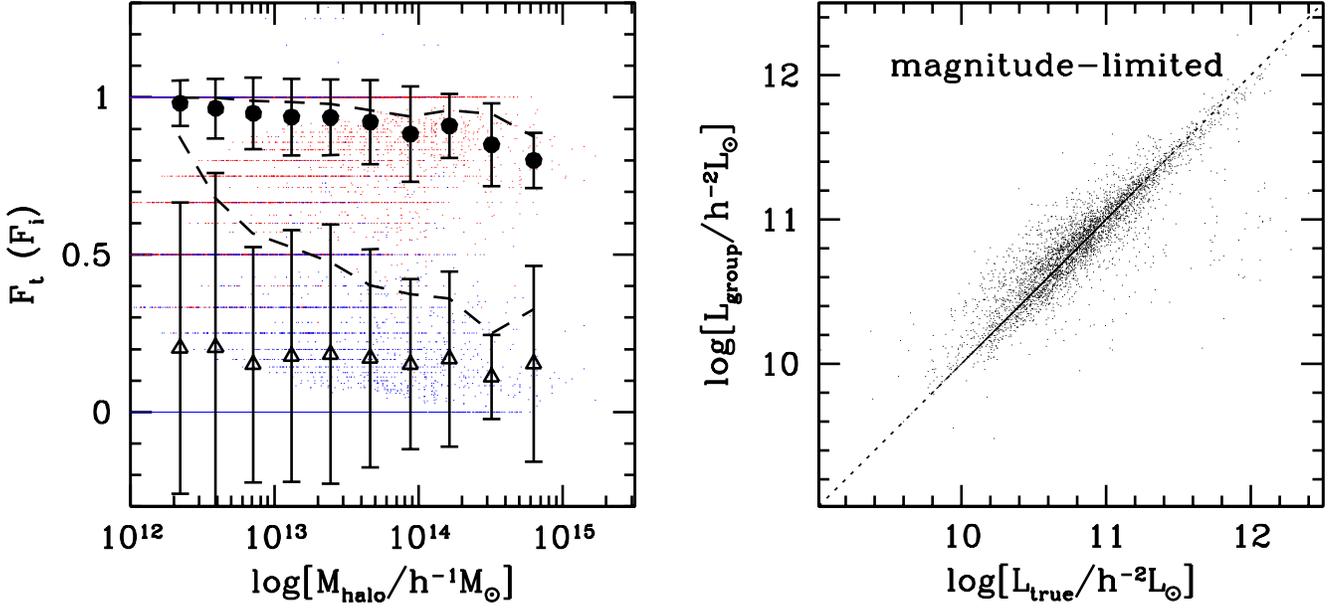,width=\hdsize}}
\caption{The left-hand panel shows the completeness ($F_{\rm t}$) and 
  contamination($F_{\rm  i}$) of the  groups selected  from our  MGRS. 
  Symbols  are as  in Fig.~\ref{fig:complete},  with the  dashed lines
  indicating  $F_{\rm t}$  and  $F_{\rm i}$  obtained  when using  the
  traditional    FOF   method    with   the    linking    lengths   of
  eq.~(\ref{eq_link1})  and~(\ref{eq_link2}).   Note  how this  method
  yields  much  larger interloper  fractions  than  our group  finder,
  especially in low-mass halos.  The right-hand panel plots the group
  luminosity, $L_{\rm group}$, for groups with $N > 2$ 
  obtained  using our fiducial  group finder,
  versus   the   true   group   luminosity   $L_{\rm   true}$   (cf.   
  Fig~\ref{fig:L_L}).}
\label{fig:2dFm_com_L}
\end{figure*}
\begin{figure*}
\centerline{\psfig{figure=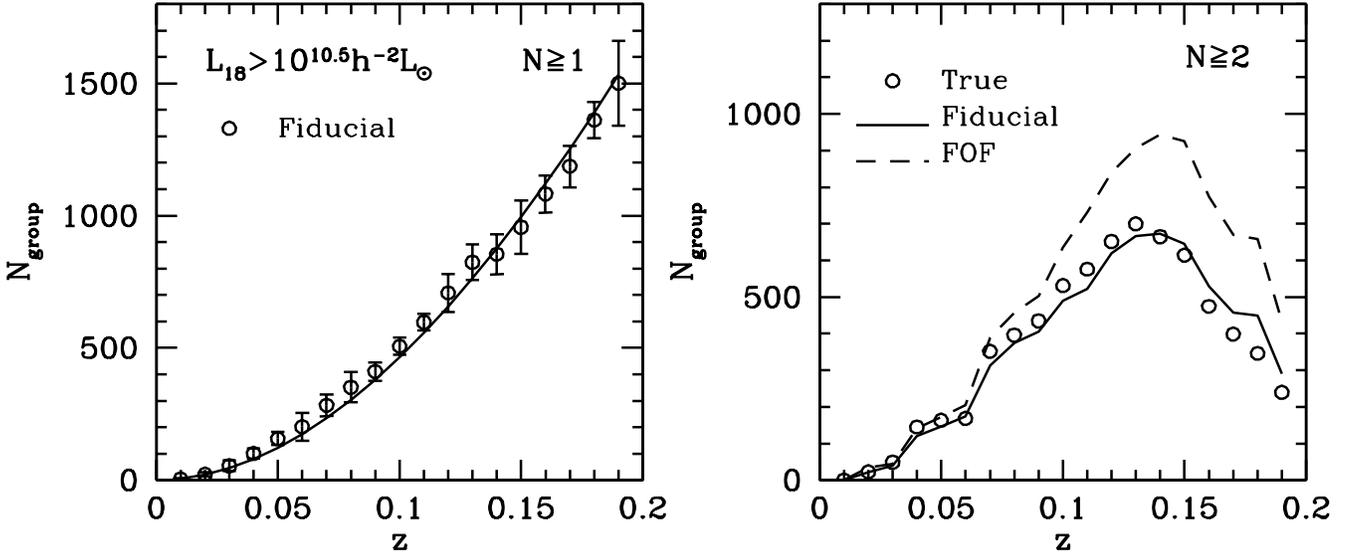,width=\hdsize}}
\caption{The number of groups detected in our MGRS as function of 
  redshift. In the left-hand panel we plot results for all groups with
  $N \geq  1$ and $L_{18}  > 10^{10.5}h^{-2}L_{\odot}$, with  the open
  circles indicating  the mean number  of groups identified  using our
  fiducial  group  finder.   The  errorbars  indicate  the  1-$\sigma$
  variance obtained  from all  8 MGRSs. The  solid line  indicates the
  expected relation for  a constant group number density  (i.e., for a
  group detection  completeness that is redshift  independent), and is
  shown  for comparison.   In the  right-hand panel  we show  the same
  $N_{\rm group}(z)$  but this time only  for groups with $N  \geq 2$. 
  The  open circles  indicate the  true $N_{\rm  group}(z)$,  with the
  solid  and dashed lines  indicating the  numbers of  groups detected
  using  our fiducial  group finder  and the  traditional  FOF method,
  respectively.}
\label{fig:N_z}
\end{figure*}
\begin{figure*}
\centerline{\psfig{figure=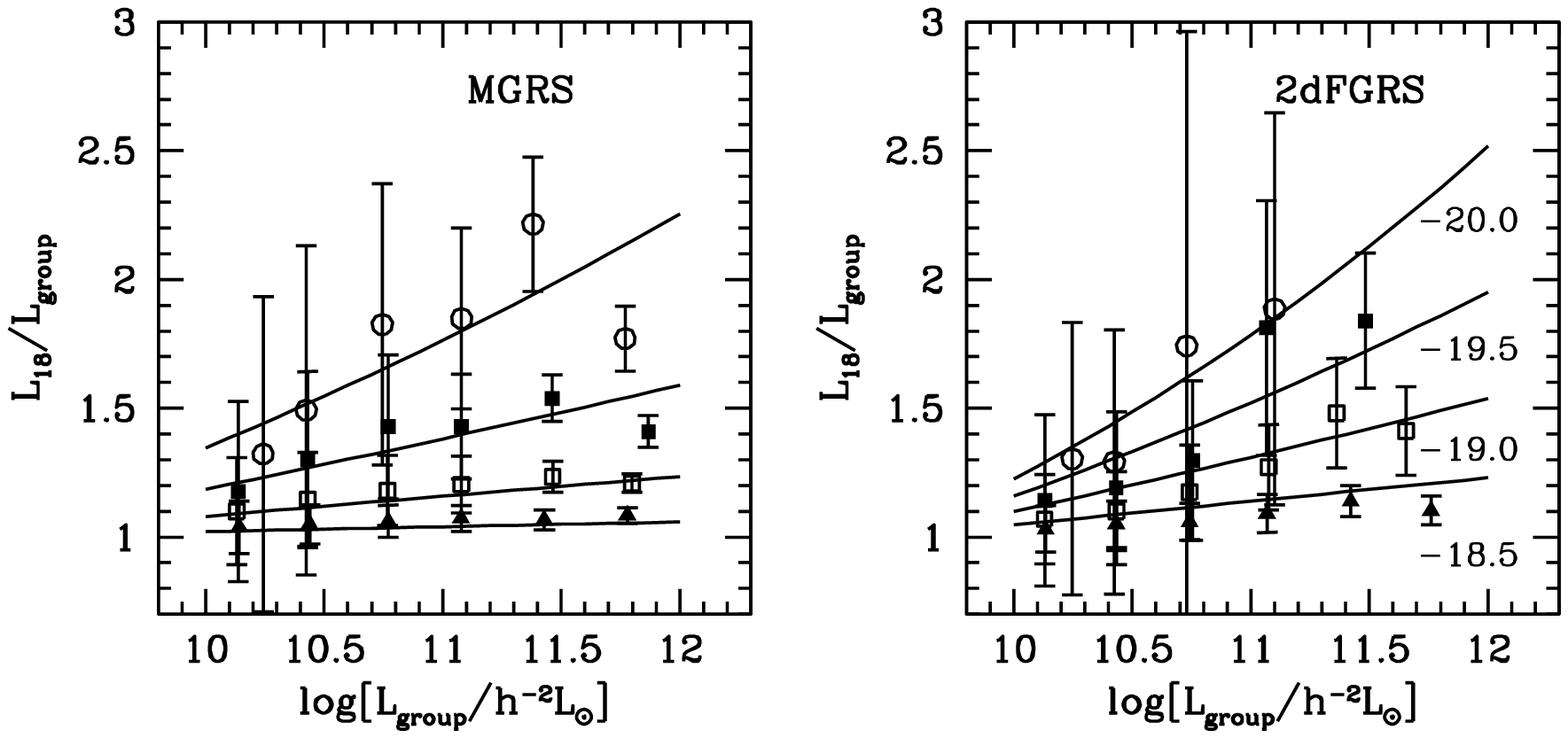,width=\hdsize}}
\caption{The ratio $L_{18}/L_{\rm group}$ as function of $L_{\rm
    group}$.   Here $L_{18}$  is  the total  luminosity  of all  group
  galaxies brighter than  $M_{b_J} - 5 \log h  = -18.0$, while $L_{\rm
    group}$ is defined  as the total luminosity of  all group galaxies
  with $M_{b_J} - 5\log h  \leq -20.0$ (open circles), $-19.5$ (filled
  squares), $-19.0$  (open squares),  and $-18.5$ (filled  triangles). 
  The  errorbars  indicate 1-$\sigma$  scatter  of  the ratios  within
  different group luminosity bins, while  the solid lines are our fits
  to these ratios,  used to compute $L_{18}$ from  an observed $L_{\rm
    group}$ (see text for details).  Results in the panels on the left
  and  right correspond  to  groups  identified in  our  MGRS and  the
  2dFGRS, respectively.}
\label{fig:LvzL18}
\end{figure*}

Having tested our group finder  on volume limited mock samples, we now
turn to more  realistic, flux-limited samples.  To that  extent we use
our  mock  2dFGRS  (hereafter  MGRS),  the construction  of  which  is
described in Section~\ref{sec:construct} and,  in more detail, in Yang
\etal (2004).

Applying the group  finder to these MGRSs yields  the completeness and
contamination     shown     in      the     left-hand     panel     of
Fig.~\ref{fig:2dFm_com_L}  (symbols with  errorbars). The  results are
very similar  to those  for our volume  limited mock samples  shown in
Fig.~\ref{fig:complete}, indicating that our group finder works almost
equally well for  flux limited samples as for  volume limited samples. 
For  comparison,  the  dashed  lines  indicate  the  completeness  and
contamination  obtained using  the standard  FOF group  finder  of Eke
\etal (2004a), with the  linking lengths given by eq.~(\ref{eq_link1})
and~(\ref{eq_link2}). Clearly, this  group finder yields significantly
more  interlopers than our  method, especially  in small  groups.  The
right-hand panel  of Fig,~\ref{fig:2dFm_com_L} compares  the estimated
group luminosity  (obtained using our fiducial group  finder) with the
true group  luminosity. Again  the results are  very similar  to those
obtained  using volume  limited samples  (cf.~Fig.~\ref{fig:L_L}), and
indicate that group luminosities can be obtained fairly robustly. 
The scatter in ${\rm log}L$ is about 0.1 in both directions,
corresponding to an error of $\sim 30\%$ when the luminosity of a 
selected group is used to infer the true luminosity. 

As  a final  demonstration of  the  performance of  our group  finder,
Fig.~\ref{fig:N_z} plots the number of groups identified as a function
of redshift. In the left-hand panel, we plot the redshift distribution
of all identified  groups (including those with only  one member) with
luminosity  $L_{18}>10^{10.5}h^{-2}L_{\odot}$  (where  $L_{18}$  is  a
scaled group luminosity which  will be defined later). Comparing these
result with  the solid  line, which corresponds  to a  constant number
density, indicates  that the  group finder works  remarkably uniformly
over  the entire  redshift  range probed:  the  group completeness  is
virtually independent of redshift.  This  success is partly due to the
fact that our group finder can also identify systems that contain only
one  or two galaxies.   If we  impose a  richness threshold,  then the
number of groups  will decline with redshift at high  $z$, as shown in
the right-hand  panel of Fig.~\ref{fig:N_z}. However,  the change with
redshift  of the number  of the  groups selected  by our  group finder
matches that of the true groups remarkably well.  In contrast, the FOF
method of Eke \etal (2004a)  yields far too many groups.  For example,
for  groups with  $L_{18}>10^{10.5}h^{-2}L_{\odot}$, our  group finder
selects 7091 groups  with richness $N\ge 2$, close  to the true number
of 7040, while the FOF method  selects 9665 groups.  For $N\ge 3$, the
true number of groups is 5231, while the number of groups selected are
4889  by our method  and 6786  by the  FOF method;  for $N\ge  4$, the
corresponding numbers are 3820, 3567, and 4969.

Although  the completeness  of the  group identification  is virtually
redshift  independent, the individual  completeness of  the identified
groups does depend on redshift. Since in an apparent magnitude limited
sample the  mean number density of galaxies  decreases with increasing
redshift,  groups associated  with similar  halos will  be  richer at
lower  redshift.   Caution is  therefore  required  when using  groups
selected  from  an  apparent-magnitude  limited sample  to  study  the
intrinsic group  properties.  For example, if we  consider groups with
similar richness,  we may in fact  mix systems with  different masses. 
One  can circumvent this  problem by  using volume-limited  samples to
identify groups. This has the advantage that the identified groups are
uniform in redshift, but the  disadvantage that it only makes limited
use of the  full observational data set.  

An alternative  method, which we  adopt in this  paper, is to  use the
full apparent-magnitude  limited sample to identify  galaxy groups. In
this case,  however, before  comparing groups at  different redshifts,
one need  to bring their  intrinsic properties as determined  from the
detected  member galaxies to  a common  scale.  Here  we focus  on the
group luminosity  and investigate how  to scale $L_{\rm  group}$.  One
possibility     is     to     compute    $L_{\rm     total}$     using
eq.~(\ref{eq:L_total}). In fact, many  earlier analyses have used this
approach to   calculate the total luminosity or richness of the groups
(Tucker 2000;     Merch\'an \& Zandivarez 2002;   Kochanek \etal 2003; 
Eke \etal 2004b).   This   method is based on the assumption that  the 
galaxy luminosity function   in groups  is  similar to  that  of field  
galaxies.   However,   as  shown in  our earlier analyses,  the galaxy  
luminosity function in  different halos     (Yang, Mo \& van den Bosch 
2003; van den Bosch, Yang \& Mo 2003)   and  different
environments (Mo \etal 2004) may  be very different.  Therefore, it is
not reliable to use  eq~(\ref{eq:L_total}) to estimate $L_{\rm total}$
from $L_{\rm  group}$.  Here we  suggest a more empirical  approach. A
nearby group  selected in an apparent magnitude  limited survey should
contain  all  of its  members  down to  a  faint  luminosity.  We  can
therefore use  these nearby groups  to determine the  relation between
the  group luminosity  obtained  using only  galaxies  above a  bright
luminosity  limit and  that obtained  using galaxies  above  a fainter
luminosity     limit.     Assuming     that    this     relation    is
redshift-independent,  one can  correct the  luminosity of  a high-$z$
group,  where   only  the  brightest  members  are   observed,  to  an
empirically normalised luminosity scale. 

As common luminosity scale we  use $L_{18}$, defined as the luminosity
of all group members brighter than $M_{b_J}=-18+5\log h$. To calibrate
the relation between $L_{\rm group}$  and $L_{18}$ we first select all
groups with $z \leq 0.09$, which corresponds to the redshift for which
a galaxy with $M_{b_J}=-18+5\log h$  has an apparent magnitude that is
equal  to   the  mean  limit  of   the  2dFGRS  ($b_J   \leq  19.3$).  
Fig.~\ref{fig:LvzL18} plots the  ratio $L_{18}/L_{\rm group}$ obtained
from  these  groups, where  $L_{\rm  group}$  is  now defined  as  the
luminosity of all group members brighter than a given limit. Different
symbols correspond to different luminosity limits, as indicated in the
right-hand  panel. The left  and right-hand  panels correspond  to our
MGRS  and the  2dFGRS, respectively.  It is clear that, for a variety  
of luminosity limits, the relation is quite tight. 

In what follows we proceed with two different ways. 
For all selected groups with $z
\leq 0.09$ we compute $L_{18}$ directly from the selected members with
$M_{b_J} \leq -18+5\log h$. For groups at higher redshifts, we compute
$L_{\rm  group}$ and  use  the appropriate,  average relation  between
$L_{18}$  and   $L_{\rm  group}$  to   estimate  the  former.   

\begin{figure}
\centerline{\psfig{figure=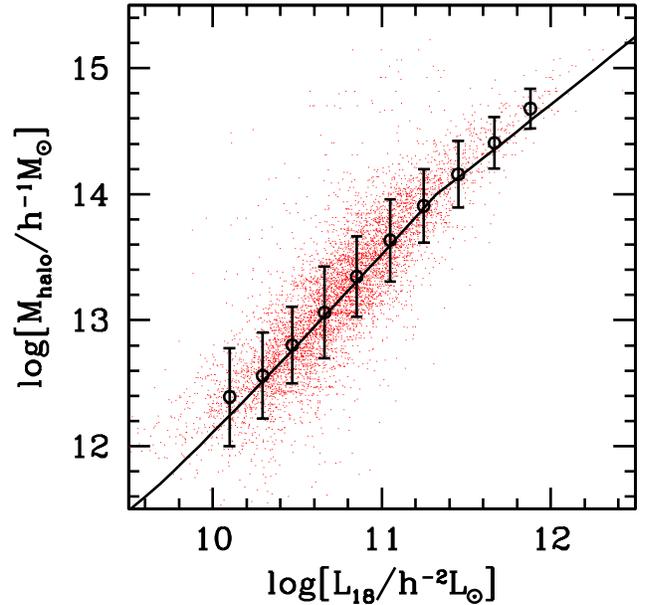,width=\hssize}}
\caption{The luminosity $L_{18}$ (see text for definition) of groups
  selected  from  the MGRS  versus  the  associated  halo mass.   Open
  circles with errorbars indicate  the mean and 1-$\sigma$ variance of
  the  distribution of halo  mass for  groups with  constant $L_{18}$,
  while the solid line corresponds to the prediction computed from the
  CLF.}
\label{fig:L18Mh}
\end{figure}
\begin{figure*}
\centerline{\psfig{figure=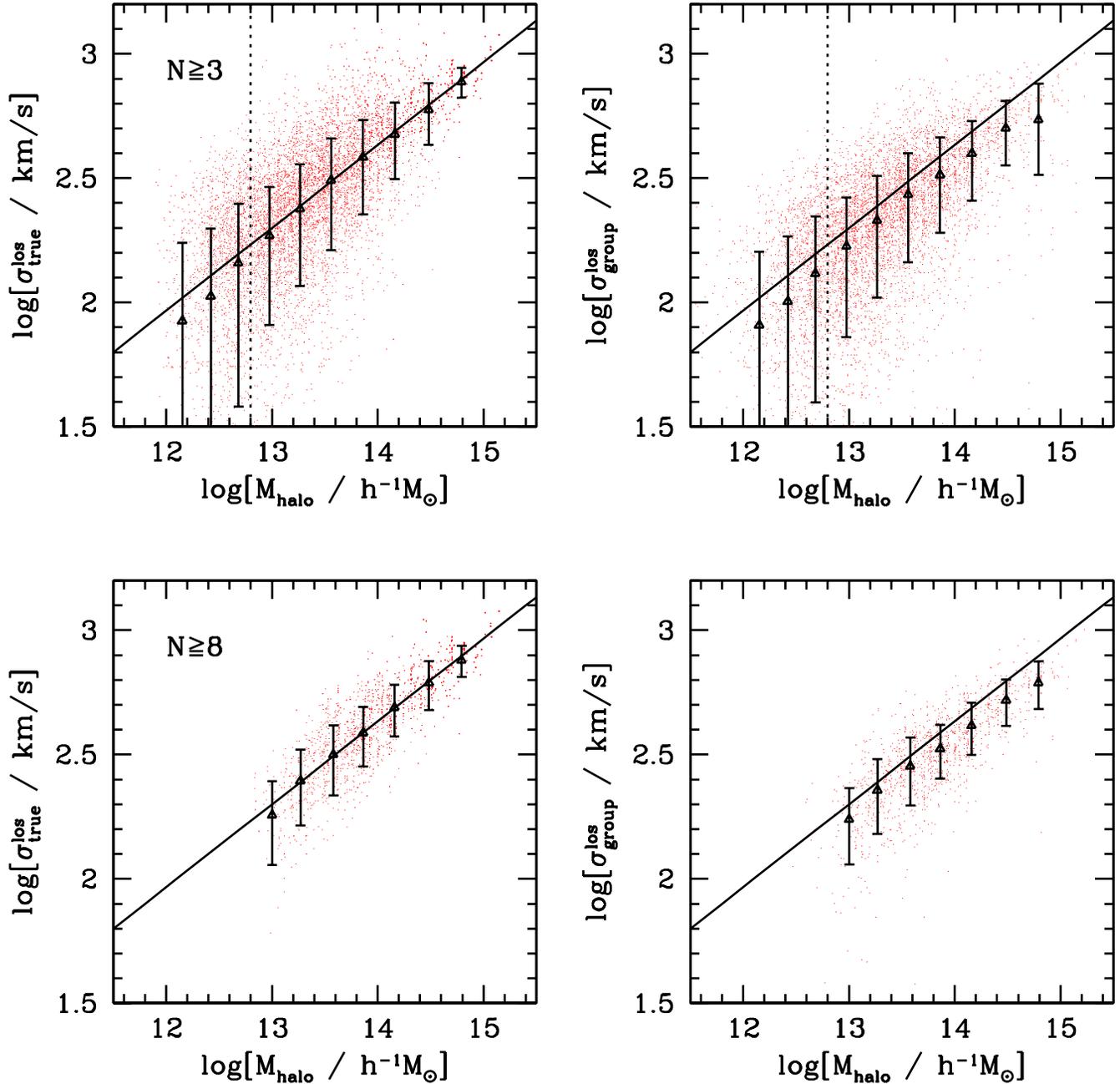,width=\hdsize}}
\caption{The left panels show the relation between the line-of-sight
  velocity dispersion of (all) true members versus the halo mass in an
  apparent magnitude  limited sample.  The right panels  show that for
  groups with identified members (true plus interloper). The upper and
  lower panels correspond to results  of groups with richness at least
  3 and 8,  respectively.  The symbols and errorbars  are the mean and
  1-$\sigma$ scatter of the  line-of-sight velocity dispersion in each
  mass bin.  The solid line  is the mass-velocity relation of the dark
  matter particles  within halos  and $M\propto \sigma^3$.  The dotted
  lines in the two upper panels roughly correspond to the completeness
  limit of the groups with members at least 3.}
\label{fig:VvV}
\end{figure*}

Fig.\,\ref{fig:L18Mh} plots $L_{18}$ thus  obtained from the groups in
our MGRS  as function of group  halo mass. Clearly,  $L_{18}$ is quite
tightly correlated with  the halo mass in the model,  and so the value
of  $L_{18}$ can  be used  to divide  groups according  to  their halo
masses. For comparison, we also plot the theoretical prediction of the
model group $L_{18}$ - halo mass relation. The excellent match between
the model prediction and the mean value confirms the accuracy of our 
group $L_{18}$ estimation.

The  velocity  dispersion of  galaxies  in  a  group is  another  mass
indicator  that  can  be  estimated  directly.  As  shown  in  Section
\ref{sec:test2},  the velocity  dispersion obtained  directly  from the
identified members  (true plus interlopers)  may differ systematically
from the  true velocity dispersion  (obtained from all true  members). 
Moreover,  in  the  2dFGRS,  the  measurement  error  in  redshift  is
typically about  $85\kms$, which is  quite big in comparison  with the
typical velocity  dispersion of  small groups.  Therefore,  the masses
estimated from velocity  dispersions are not expected to  be accurate. 
In the upper  two panels of Fig.\,\ref{fig:VvV}, we  show the relation
between the  line-of-sight velocity  dispersion of a  group (estimated
using  equation \ref{eq_gapper})  and the  mass of  the host  halo for
systems which  contain at least 3  members.  The left  panel shows the
result where all  true member galaxies in a halo  are used to estimate
the velocity dispersion. Even in this case, the scatter is quite large,
although there is no noticeable deviation of the mean from the expected
relation (solid  lines).  The right  panel shows the result  where all
selected members  (selected true members  plus interlopers) are  used. 
In   this  case,   the  presence   of  interlopers   contaminates  the
mass-velocity  dispersion  relation, and  the  velocity dispersion  is
slightly under-estimated  with respect to the expected  value.  In the
lower panels  of Fig.\,\ref{fig:VvV}, we plot the  same relations, but
for groups with at least 8  members.  Here the scatter in the relation
is greatly reduced.  In the case  where all true members are used, the
velocity dispersion of galaxies in a  group is a good indicator of mass
(see  the lower  left panel),  but bias  still exists  when  using the
selected group  members (lower right-hand  panel). This is due  to the
fact that member galaxies with  the highest velocities are the easiest
to miss by the group finder.

\begin{figure*}
\centerline{\psfig{figure=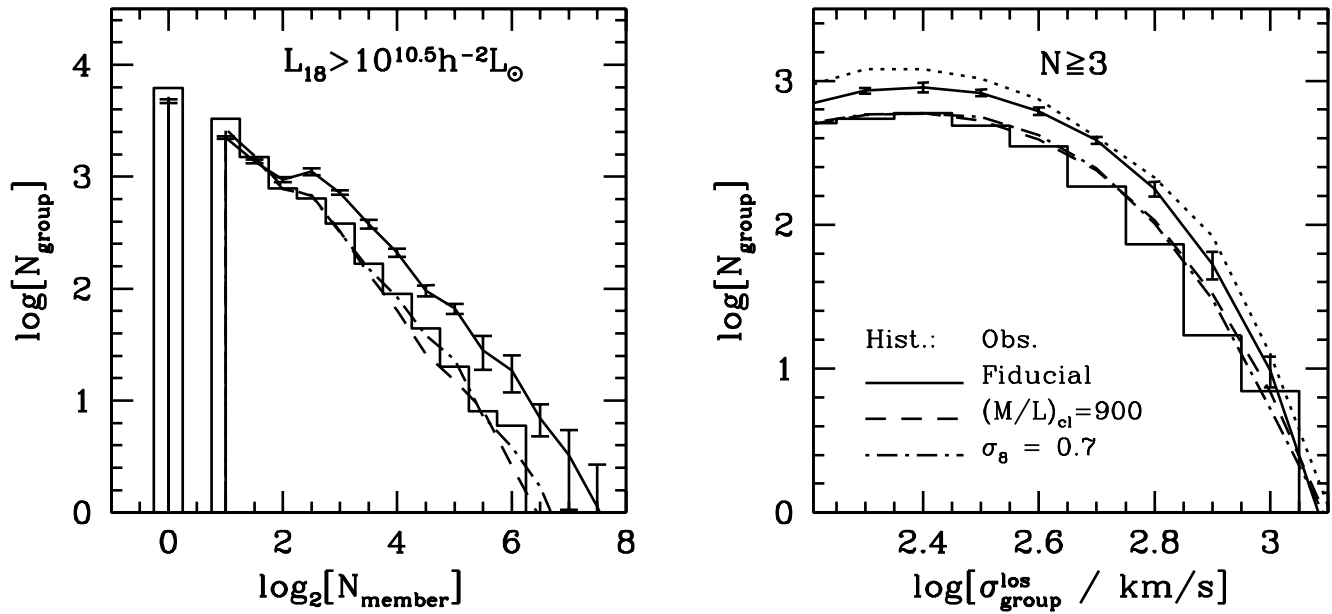,width=\hdsize}}
\caption{The richness (left panel) and group line-of-sight 
  velocity  dispersions (right panel)  of the  groups found  in 2dFGRS
  observations and mock samples.  The histograms are results of 2dFGRS
  observations, and  different lines  are results of  different model
  mock  samples. The  errorbars  are 1-$\sigma$  variances  of 8  mock
  samples. Obviously, there are more rich systems in the fiducial mock
  2dFGRS samples than in the observations. As a comparison, the dotted 
  line in the right panel shows the velocity dispersion of dark matter 
  particles within the host halos.
  }
\label{fig:2dFmo_N_z}
\end{figure*}
\begin{figure*}
\centerline{\psfig{figure=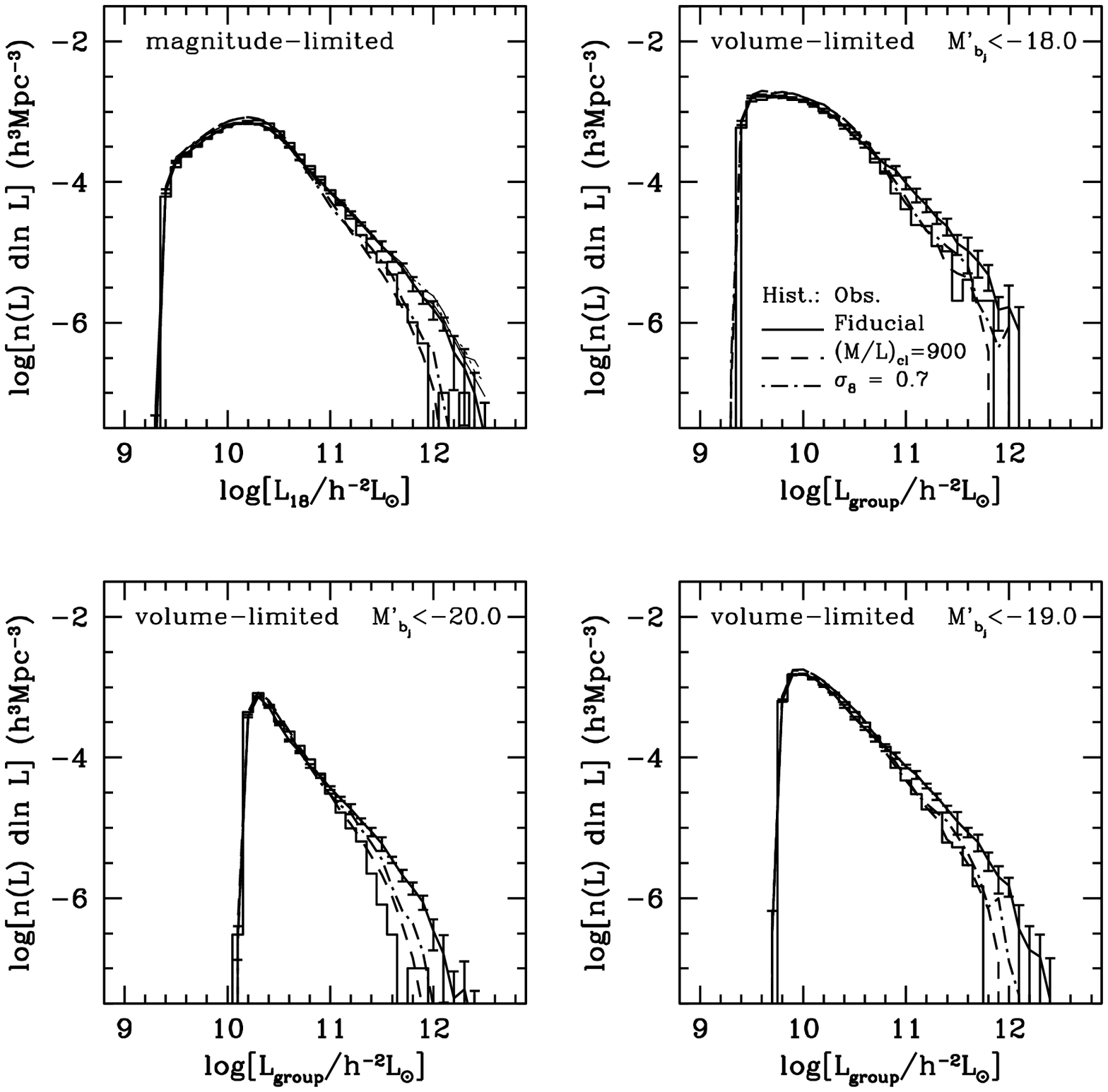,width=\hdsize}}
\caption{Similar to Fig.~\ref{fig:LF}, but for 2dFGRS observations and
  mock samples.  The histograms  are results of  2dFGRS observations,
  and different  lines are  results of different  model mock  samples. 
  The errorbars are 1-$\sigma$ variances of 8 mock samples.  The upper
  left  panel shows  the group  luminosity function for magnitude 
  limited   2dF samples. The  upper right,  lower right  and  lower  
  left  panels are  the  group   luminosity  functions  for  volume  
  limited 2dF  samples  with  absolute   magnitude  cut,  
  $M_{b_j}-5\log  h  <  -18.0,  -19.0$  and  $-20.0$,  respectively.  
  Again, we find more  bright (rich) groups in the fiducial mock
  2dFGRS samples than in the observations.  To test the impact
  of satellite galaxy distribution in individual halos, 
  we also plot in the upper left panel the results for the MGRSs that 
  assume velocity bias (thin solid line), spatial bias 
  (thin dotted line), and scatter in the $M/L$ (thin long dashed line).  
  Details about these tests are described in the text.}
\label{fig:2dFmo_LF}
\end{figure*}

\section{Applications to the 2dFGRS}
\label{sec_2dF}

Having calibrated and  tested our group finder to  both volume limited
mock samples and to flux  limited mock galaxy redshift surveys, we now
apply it  to the 2dFGRS. We  use the final public  data release, which
contains about $250,000$ galaxies with redshifts and is complete to an
extinction-corrected  apparent magnitude  of $b_J\approx  19.45$.  The
survey covers an area of  $\sim 1500$ square degrees selected from the
extended APM Survey (Maddox \etal 1996).  The survey geometry consists
of two  separate declination strips  in the North Galactic  Pole (NGP)
and  the South Galactic  Pole (SGP),  respectively, together  with 100
2-degree fields spread randomly in the southern Galactic hemisphere.

In what follows we restrict  ourselves only to galaxies with redshifts
$0.01 \leq z \leq 0.20$ in  the NGP and SGP subsamples with a redshift
quality parameter $q \geq 3$ and a redshift completeness $>0.8$.  This
leaves a grand total of $151,280$ galaxies with a typical rms redshift
error  of $85  \kms$ (Colless  \etal 2001).   Absolute  magnitudes for
galaxies  in   the  2dFGRS  are  computed   using  the  type-dependent
K-correction of  Madgwick \etal (2002).  For those  galaxies without a
reliable  estimate of $\eta$  we adopt  the average  K-correction (see
Madgwick  \etal   2002  for  details).    We  use  exactly   the  same
restrictions  when selecting galaxies  from each  of our  eight MGRSs,
yielding samples with $152,000 \pm 4,000$ mock galaxies.

\subsection{Groups in the 2dFGRS} 

Applying our fiducial group finder to our 2dFGRS sample yields a group
catalogue  containing 78708  systems. Among  these are  7251 binaries,
2343 triplets,  and 2502 groups with  four members or  more. Using the
FOF  group finder,  Eke \etal  (2004a) identified  $7020$  groups with
richness $N \geq  4$. Therefore, our
group finder seems to give a much smaller number of $N \geq 4$ systems
than does the FOF method adopted  by Eke \etal  There are two reasons
for   this    discrepancy.    First   of   all,    as   discussed   in
Section~\ref{sec:test_mag}, the  FOF method  tends to select  too many
groups  at high $z  \gta 0.1$.  Secondly, our  sample is  about $25\%$
smaller than theirs, because we have  imposed a more strict cut in sky
coverage to  get rid of regions where  observational incompleteness is
relatively high.  When applying the FOF group finder adopted by Eke 
\etal to our 2dFGRS sample, we find 5749 groups with richness 
$N \geq  4$, which is consistent with the $25\%$ difference in sample
size.  Taking these two effects into account, the number of
groups we obtain is roughly comparable to that obtained by Eke et al. 
Of  course, since  we use  a different  group finder,  the  systems we
select are  expected to  be different  from theirs, and  so we  do not
expect to have an exact match in group number.

The histograms  in Fig.~\ref{fig:2dFmo_N_z} show the  number of groups
selected from  the 2dFGRS as a  function of richness  (left panel) and
line-of-sight  velocity dispersion  (right panel).   In  the left-hand
panel  we  only  show  results  for groups  with  luminosity  $L_{18}>
10^{10.5}  h^{-2}L_{\odot}$, which  approximately  corresponds to  the
completeness   limit   of   the   group   luminosity   function   (see
Fig~\ref{fig:LF}), while  the right-hand panel only  shows results for
groups with $N \geq 3$.   The various curves in both panels correspond
to model  predictions obtained from various mock  catalogues, which will
be discussed in the next subsection. As a comparison, we plot
in the right panel of  Fig.~\ref{fig:2dFmo_N_z} 
the velocity dispersion function (which is equivalent to the mass function) 
of the dark matter particles within the mock halos. There is 
significant difference between the group velocity dispersion 
function and the halo velocity dispersion function, 
which is due to the bias already shown in Fig~\ref{fig:VvV}
and discussed in Section~\ref{sec:test_mag}. 

The   histograms  in   Fig.~\ref{fig:2dFmo_LF}  show   the  luminosity
functions of  the 2dFGRS groups.  The  upper left panel  is the result
for  the full  magnitude-limited sample.  In this  case,  the observed
group luminosities are corrected to $L_{18}$ using the relations given
in Fig.~\ref{fig:LvzL18}.  The other  three panels are the results for
volume-limited  samples.   Comparing  the group  luminosity  functions
(GLFs) for  the different volume-limited  samples, we see  that groups
with  more  than  3  members  are complete  to  a  luminosity  $L_{\rm
  group}\sim  10^{11}h^{-2}L_{\odot}$  for  the volume-limited  sample
with  $M_{b_J}<-20.0+5{\rm   log}h$,  and  to   a  luminosity  $L_{\rm
  group}\sim  10^{10.5}h^{-2}L_{\odot}$ for the  volume-limited sample
with $M_{b_J}<-18.0+5{\rm log}h $.

\subsection{Comparison between model and observation}

The results obtained above for  the 2dFGRS groups can be compared with
model predictions to constrain theories of structure formation.  Since
our MGRSs  have been constructed to  mimick the 2dFGRS  in detail, and
since we have  applied exactly the same selection  criteria as for the
2dFGRS, we  can make such a  comparison in a  straightforward way.  

The  solid curves in  Figs.~\ref{fig:2dFmo_N_z} and~\ref{fig:2dFmo_LF}
show the predictions obtained  from our MGRSs.  The errorbars indicate
the scatter from among our eight independent MGRSs, and illustrate the
expected  scatter  due  to  cosmic  variance. There  is  a  pronounced
discrepancy between these MGRSs and the 2dFGRS: the model predicts too
many  rich  systems  that  have  high velocity  dispersions  and  high
luminosities.  This  is consistent with  Yang \etal (2004),  who found
that the same mock galaxy redshift surveys predict too high amplitudes
for the pairwise  peculiar velocity dispersion and for  the real space
two-point  correlation function  on small  scales.  As  shown  in Yang
\etal, these  discrepancies can  be attributed to  the fact  that this
model predicts too many galaxies  in massive clusters. 

Before preceeding to seek potential solutions to the problem,
we first test whether the assumption of satellite galaxy 
distribution in individual halos have signifiant impact 
on our results. Firstly, we consider a case in which satellite 
galaxies have a velocity bias relative tot he dark matter.
As in Yang et al. (2004), we assume the velocity bias parameter
$b=\sigma_{\rm gal}/\sigma_{\rm DM}=0.6$. As an example, 
we show the result of the group luminosity function given by
this model as the thin solid line in the upper left panel of 
Fig.~\ref{fig:2dFmo_LF}. Secondly, we consider a case where 
the distribution of satellite galaxies in a halo is assumed to be 
shallower than that of the dark matter. Note that this assumption
is in fact consistent with the result obtained by van 
den Bosch \etal (2004b). As a simple model, we assume 
the number density distribution of satellites in a halo
to be  
\begin{equation}
\label{nsatr}
n_{\rm sat}(r) \propto \left( {r \over {\cal R} r_s} \right)^{-\alpha}
\left( 1 + {r \over {\cal R} r_s} \right)^{\alpha-3}\,,
\end{equation}
with $\alpha=0.0$ and ${\cal R}=2.0$. The group luminosity 
function predicted by this model is plotted as the thin dotted line 
in the upper left panel of Fig.~\ref{fig:2dFmo_LF}. 
Thirdly, we consider a case where we introduce
20\% scatter around the mean $M/L$ predicted by the CLF when 
populating dark matter halos with galaxies.  
Note that even without this scatter, there is scatter 
in the $M/L$ among halos of the same mass, because 
satellite galaxies are assigned to individual halos in a 
stochatical way (see Yang et al. 2004). The group luminsity 
function predicted by this model galaxies is shown as the thin long 
dashed line in  the upper left panel of Fig.~\ref{fig:2dFmo_LF}. 
Obviously, all these changes have only negligible
impact on our final results. We can therefore rule out 
the possibility that the discrepancy between our fiducial model
and observation is due to our assumption about the distribution 
of satellite galaxies in individual halos. 

Next, we seek potential solutions to the discrepancy 
between our MGRSs and the 2dFGRS.
The CLF used to construct these MGRSs was constrained to yield
an average  mass-to-light ratio for clusters of  $(M/L)_{\rm cl}=500 h
\MLsun$.   Either  the  true  mass-to-light  ratios  of  clusters  are
significantly higher (i.e., fewer  galaxies per cluster), or there are
fewer massive halos, which  implies a reduction of the power-spectrum
normalisation $\sigma_8$. Yang \etal (2004) were able to reproduce the
clustering  properties  of  2dFGRS  galaxies  if  either  the  cluster
mass-to-light  ratio is  increased to  about $(M/L)_{\rm  cl} =  900 h
\MLsun$, or $\sigma_8$ is lowered to about $0.7$ (compared to $0.9$ in
the concordance cosmology adopted  so far).  The dashed and dot-dashed
curves  in Figs.~\ref{fig:2dFmo_N_z}  and~\ref{fig:2dFmo_LF}  show the
group properties  obtained using  MGRSs with $(M/L)_{\rm  cl} =  900 h
\MLsun$ and $\sigma_8=0.7$\footnote{
 As in Yang \etal (2004), the model results for $\sigma_8=0.7$ 
 are not based on new simulations, but on the re-scaling of the 
 halo number density from the simulations of the $\sigma_8=0.9$ model.}, 
respectively. Both very nicely match the
multiplicity  function,  the  velocity  dispersion function,  and  the
luminosity  function  of the  2dFGRS  groups,  thus strengthening  the
conclusions  reached by Yang  \etal (2004)  that either  clusters have
high mass-to-light  ratios, which seems incompatible with  a number of
independent  observational constraints  (e.g., Carlberg  et  al. 1996;
Fukugita  et  al  1998;  Bahcall  2000), or  that  the  power-spectrum
normalisation  is  significantly  lower  than typically  assumed.   As
discussed  in van  den Bosch,  Mo \&  Yang (2003)  this  latter option
cannot be ruled out by current observations.

\section{Conclusions}
\label{sec_conclusion} 

The current  CDM scenario,  in which galaxies  are assumed to  form in
dark matter halos, is extremely successful in explaining a large range
of observational  data. Motivated  by this success,  and aided  by our
detailed  knowledge of the  abundances and  properties of  dark matter
halos within this CDM scenario, we have developed a group finder that
can successfully group galaxies in redshift surveys according to their
common  halos.   Using detailed  mock  catalogues, constructed  using
large numerical  simulations combined with  the conditional luminosity
function of galaxies, we carefully tested the performance of this group
finder.   Individual groups selected  using our  group finder  have an
average  completeness of  about 90  percent  and with  only $\sim  20$
percent interlopers. The group luminosities agree with the true values
to better than $70\%$ level, and the overall group luminosity function
matches the real one well for groups with $L_{\rm group}\gta 10^{10.5}
h^{-2} \Lsun$.

We  have applied  our  group finder  to  the 2dFGRS  and compared  the
properties of the  2dF groups with those extracted  from detailed mock
galaxy  redshift  surveys.   Although  the  2dF  groups  have  similar
properties as  the mock  groups, we find  a clear  discrepancy between
mock and  2dF groups, in  the sense that  the model predicts  too many
rich systems.  In order to match the observational results, we have to
either increase the  mass-to-light ratios of rich clusters  to a level
significanly higher  than the  typical observational value,  or assume
that  $\sigma_8 \simeq  0.7$ compared  to the  `concordance'  value of
$0.9$.  This result is in perfect agreement with our previous findings
based on the redshift-space  clustering of galaxies (Yang \etal 2004),
and enforces  the conclusion  that the concordance  $\Lambda$CDM model
may have too high clustering power on small scales.

The groups identified by  our  group  finder are closely related  
to the underlying  dark matter halos. Given a uniform group  
catalogue constructed in this way, 
one can do many interesting  things  that can provide pivotal 
information about  how galaxies form and evolve in CDM halos. We plan  
to  come  back to  some  of  the problems in forthcoming papers.

%%%%%%%%%%%%%%%
% Acknowledgements
%%%%%%%%%%%%%%%
                                                                               
\section*{Acknowledgement}

Numerical  simulations used  in this  paper  were carried  out at  the
Astronomical Data Analysis Center  (ADAC) of the National Astronomical
Observatory, Japan. We thank the 2dF team for making their data publicly 
available.
                                                                               
%%%%%%%%%%%%%%%
% Bibliography
%%%%%%%%%%%%%%%

\label{lastpage}

\end{document}